\documentclass[epjST]{svjour}
\usepackage{amsmath,amssymb}
\usepackage{graphicx}

\newcommand{\be}{\begin{equation}}
\newcommand{\ee}{\end{equation}}
\newcommand{\ba}{\begin{eqnarray}}
\newcommand{\ea}{\end{eqnarray}}
\newcommand{\ban}{\begin{eqnarray*}}
\newcommand{\ean}{\end{eqnarray*}}
\newcommand \nn {\nonumber}

\begin{document}

\title{Production of light nuclei at colliders \\ - coalescence vs. thermal model}

\author{Stanis\l aw Mr\' owczy\' nski\footnote{e-mail: stanislaw.mrowczynski@ncbj.gov.pl}}

\institute{
Institute of Physics, Jan Kochanowski University, 
\\
ul. Uniwersytecka 7, PL-25-406 Kielce, Poland and
\\ 
National Centre for Nuclear Research,
\\ 
ul. Pasteura 7,  PL-02-093 Warsaw, Poland}

\date{November 8, 2020}

\abstract{
The production of light nuclei in relativistic heavy-ion collisions is well described by both the thermal model, where light nuclei are in equilibrium with hadrons of all species present in a fireball, and by the coalescence model, where light nuclei are formed due to final-state interactions after the fireball decays.  We present and critically discuss the two models and further on we consider two proposals to falsify one of the models. The first proposal is to measure a yield of exotic nuclide $^4{\rm Li}$ and compare it to that of $^4{\rm He}$. The ratio of yields of the nuclides is quite different in the thermal and coalescence models. The second proposal is to measure a hadron-deuteron correlation function which carries information whether a deuteron is emitted from a fireball together with all other hadrons, as assumed in the thermal model, or a deuteron is formed only after nucleons are emitted, as in the coalescence model. The $p\! -\!^3{\rm He}$ correlation function is of interest in context of both proposals: it is needed to obtain the yield of $^4{\rm Li}$ which decays into $p$ and $^3{\rm He}$, but the correlation function can also tell us about an origin of $^3{\rm He}$. 
}

\maketitle

\section{Introduction}

Production of light nuclei in nucleus-nucleus collisions has been studied for decades but plethora of experimental results from Relativistic Heavy Ion Collider (RHIC)  \cite{Adler:2001uy,Agakishiev:2011ib} and Large Hadron Collider (LHC)  \cite{Adam:2015vda,Acharya:2017fvb,Acharya:2017bso} have revived an interest in the problem and attracted a lot of attention. At high-energy collisions light nuclei occur as fragments of incoming nuclei, as at low-energy collisions, but we also deal with a genuine production process -- the energy released in a collision is converted into masses of baryons and antibaryons which form nuclei and antinuclei. 

The remnants of initial nuclei occur at rapidities of projectile and target nuclei while the genuinely produced nuclides populate a  midrapidity domain. Therefore, products of the two mechanisms are  kinematically well separated. Further on we are interested only in the midrapidity domain where numbers of the nuclei and antinuclei are approximately equal to each other at RHIC and are exactly equal at LHC. The baryon-antibaryon symmetry clearly shows that the matter created in the collisions is (almost) baryonless -- there is no net baryon charge. Together with light nuclei and antinuclei up $^4{\rm He}$ and $^4\overline{\rm He}$ there are also produced hypertritons and antihypertritons both at RHIC and LHC \cite{Abelev:2010rv,Adam:2015yta}.

According to the coalescence model \cite{Butler:1963pp,Schwarzschild:1963zz}, the production of light nuclei is a two-step process: production of nucleons and formation of nuclei due to final-state interaction among nucleons which are close phase-space neighbors. The energy scale of the first step, which is a double nucleon mass, is much bigger than that of the second one which is a nuclear binding energy. Consequently, a probability to produce a given nuclide factorizes into the probabilities to produce the nucleons and to form the nuclide. The latter probability takes into account an internal structure of the produced nucleus. 

Although the coalescence model is known to work well in a broad range of collision energies, the model is well justified when nucleons are truly produced because of the energy scale separation. So, it is not surprising that the model properly describes production of light nuclei and antinuclei at RHIC and LHC \cite{Sun:2015ulc,Sun:2017ooe,Zhu:2015voa,Zhu:2017zlb,Wang:2017smh}.

The thermodynamical model, see the review \cite{BraunMunzinger:2003zd}, is also more reliable and simpler at the highest available collision energies than at lower ones. Since thousands of hadrons are produced at colliders, it is easier to justify the statistical assumption of equipartition of energy. The model is also simpler because the matter produced at midrapidity is baryonless and consequently the baryon chemical potential vanishes. Therefore, particle's yields are determined solely by two thermodynamic parameters: the temperature and system's volume. Nevertheless the model predicts very well not only the yields of all hadron species measured at RHIC and LHC but also of light nuclei and hypernuclei \cite{Andronic:2010qu,Cleymans:2011pe,Andronic:2017pug}. The predictions depend on masses and numbers of internal degrees of freedom of the light nuclei but are independent of their internal structures. 

The evident success of the thermal model, which has attracted a lot of interest \cite{Wang:2017smh,Mrowczynski:2016xqm,Bazak:2018hgl,Mrowczynski:2019yrr,Bazak:2020wjn,Bellini:2018epz,Oliinychenko:2018ugs,Xu:2018jff,Bugaev:2018klr,Bugaev:2020sgz,Sun:2018mqq,Vovchenko:2019aoz,Vovchenko:2020dmv,Cai:2019jtk},  is very puzzling, as it is hard to expect that nuclei exist in the hot and dense fireball environment. The temperature is much bigger than a binding energy and the system is so dense that the inter-particle spacing is smaller than the typical inter-nucleon distance in a nucleus. Therefore, proponents of the thermal model speculate \cite{Andronic:2017pug} that nuclei are produced as colorless droplets of quarks and gluons with quantum numbers that match those of the final-state nuclei. 

The thermal and coalescence models are physically different but its was observed long ago that the models give rather similar yields of light nuclei \cite{DasGupta:1981xx}. The observation has been recently confirmed \cite{Zhu:2015voa,Mrowczynski:2016xqm}, using a more refined version of the coalescence model \cite{Sato:1981ez,Gyulassy:1982pe,Mrowczynski:1987,Lyuboshitz:1988,Mrowczynski:1992gc}, which properly treats a quantum-mechanical character of the formation process of light nuclei.

The aim of this short but rather pedagogical review is to critically discuss the thermal and coalescence models of production of light nuclei in relativistic heavy-ion collisions. The phenomenon has been studied experimentally and theoretically in a broad range of collision energies over decades. However, the process of a heavy-ion collision crucially depends on a collision energy. Theoretical methods which are applicable at low and high energies are also rather different. Therefore, our discussion is limited to the highest collision energies accessible at RHIC and LHC. The methods relevant at lower energies when the temperature of nuclear matter does not exceed, say, 20 MeV, are discussed in \cite{Pais:2019shp,Ropke:2020peo} and references therein. 

After presentation of the models, we discuss, following and improving the analysis of Ref.~\cite{Mrowczynski:2016xqm}, why the models predict similar yields of light nuclei. Subsequently we consider a possibility to falsify one of the models. We present two proposals. The first one is to compare the yield of $^4{\rm He}$  to that of exotic nuclide $^4{\rm Li}$ \cite{Mrowczynski:2016xqm,Bazak:2018hgl,Bazak:2020wjn}. Since the masses of both nuclei are close to each other, the yield of $^4{\rm Li}$ is according to the thermal model about five times bigger than that of $^4{\rm He}$ due to five spin states of $^4{\rm Li}$ and only one of $^4{\rm He}$. The coalescence model predicts instead a significantly smaller ratio of the yields of $^4{\rm Li}$ to $^4{\rm He}$ because the latter nuclide is well bound and compact while the former one is loose. The model also predicts that the ratio strongly depends, in contrast to the thermal model, on the collision centrality \cite{Bazak:2018hgl,Bazak:2020wjn}.

The second proposal to falsify one of the models relies on the observation \cite{Mrowczynski:2019yrr} that a hadron-deuteron correlation function can tell us whether deuterons are directly emitted from a fireball or they are formed later on due to final-state interactions. The radii of the sources of deuterons differ from each other by the factor $\sqrt{4/3}$ in the two cases. Therefore, knowing the radius of a nucleon source from the proton-proton correlation function we can quantitatively distinguish the emission of a deuteron from the fireball, as in the thermal approach, from the formation of a deuteron afterwards, as in the coalescence model. 

We also discus the $p\!-\!^3{\rm He}$ correlation function which is important, as discussed at length in \cite{Bazak:2020wjn}, in the context of both proposals. The correlation function needs to be measured to obtain the yield of $^4{\rm Li}$ which is unstable and decays into the $p\!-\!^3{\rm He}$ pair. The $p\!-\!^3{\rm He}$ correlation function also carries information analogous to that of the hadron-deuteron correlation function. If one assumes that $^3{\rm He}$ is emitted directly from the fireball the source radius inferred from the $p\!-\!^3{\rm He}$ correlation function is smaller by the factor $\sqrt{3/2}$ than that corresponding to the scenario where nucleons emitted from the fireball form the nuclide $^3{\rm He}$ due to final-state interactions.

Throughout the article we use the natural units with $c=\hbar = k_B = 1$.

\section{Coalescence and thermal models}
\label{sec-models}

Let us introduce the coalescence and thermal models. We stress again that we do not consider light nuclei which are fragments of colliding nuclei but only those genuinely produced at midrapidity in collider experiments at RHIC or LHC. 

\subsection{Coalescence model}

As already mentioned in the introduction, production of light nuclei is a two-step process in the coalescence model \cite{Butler:1963pp,Schwarzschild:1963zz}. At first nucleons are produced and later on a formation of nuclei proceeds due to final-state interactions of nucleons which are close to each other in momentum and coordinate spaces. The fact that the energy scale of the first step, which is the double nucleon mass, is much higher than that of the second one, which is a typical nuclear binding energy, is important for two reasons. First of all, the two steps of the process can be distinguished because the temporal scales are roughly inverse of the energy scales. Consequently, the production of nucleons, which occurs at first, is a fast process while the formation of nuclei is a slow process which occurs subsequently. Secondly, a probability to produce a nucleus of $A$ nucleons can be factorized into the probability to produce (independently) $A$ nucleons and the probability that nucleons fuse into the nucleus. Therefore, the number of nuclei with momentum ${\bf p}_A$ is
\be
\label{A-mom-dis}
\frac{dN_A}{d^3p_A} = {\cal A}_A \bigg(\frac{dN_N}{d^3p} \bigg)^A,
\ee
where ${\bf p}_A = A{\bf p}$ and ${\bf p}$ is assumed to be much bigger than the characteristic internal momentum of a nucleon in the nucleus of interest, ${\cal A}_A$ is the formation rate of a nucleus under consideration. 
 
One often assumes, as suggested long ago in \cite{Schwarzschild:1963zz}, that nucleons form a nucleus if they occur in a momentum sphere of a radius $p_0$. Then,
\be
\label{A-rate-pheno}
{\cal A}_A = g_S g_I \Big(\frac{4 \pi}{3} p_0^3 \Big)^{A-1} ,
\ee
where $g_S$ and $g_I$ are spin and isospin factors which take care of probability that quantum numbers of $A$ nucleons match to those of the nuclide of interest. The nucleons are assumed to be unpolarized. The momentum distributions of protons and of neutrons are assumed to be of the same shape but the numbers of protons and of neutrons can differ. The parameter $p_0$, which is roughly a momentum of a nucleon in a nucleus, is a free parameter of the model to be inferred from experimental data. 

It is also often required that nucleons, which fuse into a nucleus, must be close to each other not only in the momentum space but in the coordinate space as well, see e.g. \cite{Sombun:2018yqh}. The formula (\ref{A-rate-pheno}) is then modified. 

To formulate a relativistically covariant coalescence model one uses, see {\it e.g.} \cite{Sato:1981ez}, the Lorentz invariant momentum distributions and writes down the relation analogous to (\ref{A-mom-dis}) as
\be
\label{A-mom-dis-LC}
E_A \frac{dN_A}{d^3p_A} = B_A \bigg(E_N \frac{dN_N}{d^3p} \bigg)^A,
\ee
where $E_A$ and $E_N$ are energies of the nucleus and nucleons under consideration. Demanding that the relations (\ref{A-mom-dis}) and (\ref{A-mom-dis-LC}) are identical in the center-of-mass frame of the nucleus, which is formed, the parameter $B_A$ is found to be 
\be
\label{B-A-def}
B_A \equiv \frac{A m}{m^A} \, {\cal A}_A,
\ee
where $m$ is the nucleon mass. The form of the relation (\ref{A-mom-dis-LC}) implies that the parameter $B_A$ is a Lorentz scalar. However, one should realize that the formula (\ref{B-A-def}) does not have a solid foundation. We return to this point after a quantum-mechanical approach to the formation rate is introduced. 

The phenomenological approaches to production of light nuclei, which are based of the formula (\ref{A-rate-pheno}) or their variations, do not take into account a quantum-mechanical character of the process of a bound state formation. However, it was discovered by Sato and Yazaki  \cite{Sato:1981ez} and discussed later on by several authors, see e.g. \cite{Gyulassy:1982pe,Mrowczynski:1987,Lyuboshitz:1988}, that the formation of a nucleus driven by final-state interactions is fully analogous to the process responsible for short range correlations observed among final-state hadrons with small relative velocities. Therefore, the quantum-mechanical formula which gives the deuteron formation rate is almost identical to that of neutron-proton correlation function \cite{Mrowczynski:1992gc}. The two quantities are actually related to each other due to the completeness of quantum states \cite{Mrowczynski:1994rn,Maj:2004tb,Maj:2019hyy}.

The formation rate of a nucleus of $A$ nucleons ${\cal A}_A$ is given as
\ba 
\nn
{\cal A}_A  &=& g_S g_I (2\pi)^{3(A-1)} V
\int d^3r_1 \, d^3r_2 \dots d^3r_A \,
D ({\bf r}_1)\, D ({\bf r}_2) \dots D ({\bf r}_A)\, 
\\[2mm] \label{A-form-rate}
&& ~~~~~~~~~~~~~~~~~~~~~~~~~~~~~~~~~~~~~~~~~~~~~~~~~~~~~~
\times |\Psi({\bf r}_1,{\bf r}_2, \dots {\bf r}_A ) |^2 ,
\ea
where $g_S$ and $g_I$ are, as previously, the spin and isospin factors; the multiplier $(2\pi)^{3(A-1)}$ results from our choice of natural units where $\hbar =1$; $V$ is the normalization volume which disappears from the final formula; the source function $D ({\bf r})$ is the normalized to unity position distribution of a single nucleon at the kinetic freeze-out and $\Psi({\bf r}_1,{\bf r}_2, \dots {\bf r}_A )$ is the wave function of the nucleus of interest. 

The formula (\ref{A-mom-dis}) does not assume, as one might think, that the nucleons are emitted simultaneously. The vectors ${\bf r}_i$ with $i=1,2, \dots A$ denote the nucleon positions at the moment when the last nucleon is emitted from the fireball. For this reason, the function $D ({\bf r}_i)$ actually gives the space-time distribution. It is often chosen in the isotropic Gaussian form
\be
\label{Gauss-source}
D ({\bf r}_i) = (2 \pi R_s^2)^{-3/2} \, e^{-\frac{{\bf r}_i^2}{2R^2_s}},
\ee
where $\sqrt{3} R_s$ is the root-mean-square (RMS) radius of the nucleon source. 

The Gaussian parametrization of the source function (\ref{Gauss-source}) is not only convenient for analytical calculations but there is an empirical argument in favor of this choice. The imaging technique \cite{Brown:1997ku} allows one to infer the source function from a two-particle correlation function provided the inter-particle interaction is known. The technique applied to experimental data from relativistic heavy-ion collisions showed that non-Gaussian contributions to the source functions are rather small and do not much influence the correlation functions \cite{Alt:2008aa}.    

As already mentioned, Lorentz invariant momentum distributions are used in the relativistically covariant coalescence model and the coalescence rate formula (\ref{A-form-rate}) is modified, see e.g.  \cite{Sato:1981ez,Mrowczynski:1987}. However, the modifications are actually heuristic as a theory of strongly interacting bound states faces serious difficulties. In particular, there is no factorization of a center-of-mass and relative motion. To avoid complications one considers the formation process in the center-of-mass frame of the nucleus to be formed where the process can be treated nonrelativistically even so momenta of nucleons are relativistic in both the rest frame of the source and in the laboratory frame. The point is that the formation rate is non-negligible only for small relative momenta of the nucleons. Therefore, the relative motion can be treated as nonrelativistic and the corresponding wave function is a solution of the Schr\"odinger equation. The source function, which is usually defined in the source rest frame, needs to be transformed to the center-of-mass frame of the pair as discussed in great detail in \cite{Maj:2009ue}. 

The practical calculations of the formation rate ${\cal A}_A$ for $A=2,\,3$ and 4, which require a separation of a center-of-mass and relative motion, are presented in Secs.~\ref{sec-d/p-coalescence-model}, \ref{sec-p-3He-correlation} and \ref{sec-4Li-4He}, respectively. 

It was repeatedly stated in the literature -- starting from the very first paper on the coalescence model \cite{Butler:1963pp} -- that a neutron and proton must interact with a third body to form a deuteron as otherwise the energy and momentum cannot be conserved simultaneously. The statement, which was also extended to nuclei heavier than a deuteron, is indeed correct if the neutron and proton are both on mass shell. However, as observed long ago \cite{Mrowczynski:1987}, nucleons, which are emitted from a fireball, are {\em not} on the mass shell due to the finite space-time size of a fireball. The space-time localization of a nucleon within the fireball washes out its four-momentum due to the uncertainty principle. Using a more formal language of scattering theory, the nucleons are not in an asymptotic state in the remote past or remote future which indeed requires the mass-shell condition. Instead the nucleons are in an intermediate scattering state. Therefore, there is no reason to impose the mass-shell constraint. Because the space-time size of the fireball is of the same order as that of the nucleus, which is formed, the mismatch of the energy-momentum is washed out by the uncertainty of energy and momentum of the nucleons. 

Let us also note that the models of relativistic heavy-ion collisions like AMPT or UrQMD, which are close in spirit to Quantum Molecular Dynamics, do not treat a formation of light nuclei dynamically. Instead there are `afterburners' based on one or another version of the coalescence model, see \cite{Zhu:2017zlb} for AMPT and \cite{Sombun:2018yqh} for UrQMD. Therefore, the models do not offer another approach to the production of light nuclei.

We close the presentation of the coalescence model by saying that whenever we refer to the model we keep in mind the expression (\ref{A-mom-dis}) with the formation rate given by Eq.~(\ref{A-form-rate}). 

\subsection{Thermal model}

The fundamental postulate of the thermal model is the equipartition of fireball's energy among all degrees of freedom active in the system. Light nuclei are assumed to be populated as all other hadrons and when the fireball decays the nuclei show up in a collision final state. Their yield reflects a thermodynamic state of the fireball at the moment of chemical freeze-out when inelastic collisions of fireball's constituents become no longer operative.

The partition function is evaluated as a mixture of ideal gases of all stable hadrons and resonances. The presence of resonances corresponds to attractive interactions among hadrons. Sometimes additional repulsive interactions are modelled with an excluded volume prescription, see e.g. \cite{Bugaev:2020sgz}. However, as authors of the thermal model state \cite{Andronic:2017pug}, their results on thermal parameters remain unchanged from what is obtained in the non-interacting limit except for the overall particle density which is reduced by up to 25\%. 

In the fireball rest frame a momentum distribution of hadrons $h$ at the moment of chemical freeze-out is, see {\it e.g.} \cite{BraunMunzinger:2003zd}, 
\be
\label{mom-dis-h}
\frac{d N_h}{d^3p} = \frac{g_h \, V_{\rm chem}}{(2\pi)^3} \, e^{-\beta_{\rm chem} (E_{\bf p} - \mu)} ,
\ee
where $g_h$ is the number of internal degrees of freedom of the hadron $h$, $m_h$  is its mass and $E_{\bf p} \equiv \sqrt{m_h^2 + {\bf p}^2}$ is the energy, $V_{\rm chem}$ is the system's volume at the chemical freeze-out, $\beta_{\rm chem} \equiv 1/T_{\rm chem}$ is the inverse temperature and $\mu$ is the chemical potential related to conserved charges carried by the hadron. The baryon chemical potential usually plays an important role but, as already mentioned, the matter created in midrapidity at collider energies is baryonless and consequently the chemical potential vanishes. We note that the formula (\ref{mom-dis-h}) is classical that is it neglects effects of quantum statistics. The effects are usually minor because of many hadron species which populate many quantum states. The formula (\ref{mom-dis-h}) also neglects inter-hadron interactions. 

The yield of hadrons $h$ is
\be
\label{yield-h}
N_h = g_h V_{\rm chem} \int \frac{d^3p}{(2\pi)^3} \, e^{-\beta E_{\bf p}}  
=
\frac{g_h}{2\pi^2} \,V_{\rm chem} m_h^2 T_{\rm chem} K_2 (\beta_{\rm chem}m_h) ,
\ee
where $K_2(x)$ is the so-called Macdonald function which for $x \gg 1$ can be approximated as 
\be
\label{approx-K2}
K_2(x) =\sqrt{\frac{\pi}{2x}} \, e^{-x}  \Bigg(1  + \frac{15}{8x} + {\cal O}\bigg(\frac{1}{x^2}\bigg) \Bigg).
\ee

Experimentally observed yields of hadrons include not only the direct contribution given by Eq.~(\ref{yield-h}), but contributions, which sometimes are sizable, coming from decays of unstable states \cite{BraunMunzinger:2003zd}. Therefore, to predict a yield of, say, protons, one has to take into account all unstable hadron states which ultimately decay into a proton. The contributions are weighted with the decay branching ratios. In case of light nuclei, their yields should include nuclear excited states which decay into a nuclide of interest. 

A microscopic mechanism responsible for production of light nuclei in the fireball is unspecified and may be even unknown. As the formula (\ref{yield-h}) shows, the yields of hadrons are determined by their masses and internal degrees of freedom, and by two thermodynamical parameters: $V_{\rm chem}$ and $T_{\rm chem}$. The temperature, however, is much bigger than a typical nuclear binding energy and the inter-particle spacing is smaller than a characteristic size of light nuclei. So, the nuclei cannot exist in the fireball -- they are as `snowflakes in hell' \cite{PBM-2015}. Proponents of the thermal model argue \cite{Andronic:2017pug} that there are colorless droplets of quarks and gluons present in the fireball and those with appropriate quantum numbers are converted later on into light nuclei. 

\subsection{Do the models differ?}

The coalescence model offers a microscopic picture in a sense that it uniquely identifies the physical process responsible for light nuclei production that is the final state interaction. The thermal approach instead presents a macroscopic description. So, one wonders whether the production mechanisms of light nuclei behind the coalescence and thermal models are physically different from each other.

One can argue that instead of the two models we should rather consider, as in the study \cite{Oliinychenko:2018ugs}, hadron-hadron and hadron-deuteron interactions which are responsible for a deuteron production and disintegration in a fireball before its decay. Such an approach is physically sound if a particle source and an average inter-hadron spacing in the source are both much bigger than a deuteron size. Additionally the lifetime of the source should be much longer than the characteristic time of deuteron formation. 

However, the assumptions are rather far from reality of relativistic heavy-ion collisions. The deuteron radius is about 2 fm and the time of deuteron formation, which is of the order of the inverse binding energy, is roughly 100 fm/$c$. Consequently, the size of the particle source is of the same order as the deuteron radius, the inter-hadron spacing in the source is smaller than a deuteron, and the lifetime of the source is significantly shorter than the deuteron formation time. Therefore, a state of a neutron-proton pair in between the frequent collisions cannot be identified with an asymptotic deuteron state which is defined in a scattering theory either in a remote past or remote future. 

The coalescence mechanisms of deuteron formation and direct thermal production are physically different in relativistic heavy-ion collisions because the particle source is small and dense when compared to a deuteron and the source lifetime is shorter than the deuteron formation time. According to the coalescence model, light nuclei are formed long after nucleons are emitted from the source. The thermal model assumes that light nuclei are emitted directly from the source. 

\section{Yield of deuterons}
\label{sec-D-to-p}

As already mentioned in the introduction, its was observed long ago that the coalescence and thermal models give rather similar yields of light nuclei \cite{DasGupta:1981xx}. The results of the thermal model can be easily reproduced by means of simple formulas but those of the coalescence model are usually obtained  using Monte Carlo generators, see {\it e.g.} \cite{Zhu:2015voa}. Therefore, it is hard to see how it happens and why its happens that the models predict similar yields of light nuclei. For this reason we derived \cite{Mrowczynski:2016xqm} simple analytical formulas which give the ratio of yields of deuterons -- the simplest nuclei -- in the two models. The model parameters were inferred from experimental data. In this section we recapitulate and improve the analysis presented in \cite{Mrowczynski:2016xqm}.

\subsection{$D/p$ in thermal model}
\label{sec-D/p-thermal-model}

The yield of protons is given by the formula (\ref{yield-h}). Since $\beta_{\rm chem} m$, where $m$ is the proton mass, equals 6 for $T_{\rm chem} = 156$~MeV, we use the approximation (\ref{approx-K2}) and write the proton yield as
\be
\label{Np-chem-approx}
N_p = 2\lambda_p \,V_{\rm chem} \Big(\frac{m T_{\rm chem}}{2\pi} \Big)^{3/2} e^{- \beta_{\rm chem}m}  
\bigg(1 + \frac{15 T_{\rm chem}}{8m} + {\cal O}\Big(\frac{T_{\rm chem}^2}{m^2}\Big) \bigg) ,
\ee
where except the spin degeneracy factor 2 we have included the parameter $\lambda_p$ which takes into account a sizable contribution of protons coming from decays of baryon resonances \cite{BraunMunzinger:2003zd}. The parameter will be estimated later on. 

Since the number of deuterons is given by the formula analogous to (\ref{Np-chem-approx}), the ratio of the deuteron to proton yield is
\ba
\label{d-p-TM}
\frac{D}{p} = \frac{3 \sqrt{2}}{\lambda_p} \, e^{- \beta_{\rm chem}m} 
 \Bigg(1  - \frac{15 T_{\rm chem}}{16 m} + {\cal O}\bigg(\frac{T_{\rm chem}^2}{m^2}\bigg) \Bigg),
\ea 
where the spin degeneracy factor of a deuteron is 3 and its mass is approximated as $2m$. We note that the parameter $\lambda_D$ analogous to $\lambda_p$ is not introduced because the contribution of deuterons, which originate from decays of excited light fragments, is negligible \cite{Vovchenko:2020dmv}. 

\subsection{$D/p$ in coalescence model}
\label{sec-d/p-coalescence-model}

The momentum distribution of the final-state deuterons is given by the formulas (\ref{A-mom-dis}) and (\ref{A-form-rate}) both with $A=2$. Introducing the center-of-mass variables
\be
{\bf R} \equiv \frac{1}{2}({\bf r}_1 + {\bf r}_2) , ~~~~~~~~~~~~~~~~ {\bf r} \equiv {\bf r}_2 - {\bf r}_1,
\ee
and writing down the deuteron wave function as $\psi ({\bf r}_1,{\bf r}_2) = e^{i {\bf P}{\bf R}} \phi_D ({\bf r} )$, the deuteron formation rate equals 
\be
\label{d-form-rate}
{\cal A}_2 = \frac{3}{4} (2\pi)^3 \int d^3r \, D_r ({\bf r})\, |\phi_D ({\bf r} ) |^2 ,
\ee
where the normalized to unity `relative' source function is
\be 
\label{D-r-source-def-np}
D_r({\bf r}) \equiv \int d^3 R \, D \Big ({\bf R} + \frac{1}{2}{\bf r}\Big) 
\, D \Big ({\bf R}-\frac{1}{2} {\bf r}\Big)  = \Big({\frac{1}{4 \pi R_s^2}}\Big)^{3/2} e^{ -\frac{{\bf r}^2}{4R_{\rm kin}^2} } .
\ee
The latter equality holds for the Gaussian single-particle source function (\ref{Gauss-source}). In this section the source radius carries the index `kin' not `{\it s}' to stress that we deal with the kinetic freeze-out. The factor $\frac{3}{4}$ reflects the fact the deuterons come from the neutron-proton pairs in three spin triplet states out of four possible spin states of unpolarized two nucleons. 

To compute the deuteron yield, the nucleon momentum distribution needs to be specified. We write down the proton distribution in terms of the transverse momentum $(p_T)$,  transverse mass  $\big(m_T \equiv \sqrt{m^2 +p_T^2}\,\big)$, and rapidity $(y)$ as
\be
\frac{dN_p}{d^3p} = \frac{1}{ m_T \cosh y} \frac{dN_p}{dy\, d^2p_T},
\ee
and we choose the distribution at midrapidity which is flat in rapidity and azimuthal angle and it exponentially decays with the transverse mass that is
\be
\label{p-pT-y-dis}
 \frac{dN_p}{dy\, d^2p_T} =  \frac{N_p}{2\pi \Delta y} 
\,  \frac{e^{\beta_{\rm kin}m}}{T_{\rm kin}(m + T_{\rm kin})} \,
e^{-\beta_{\rm kin}m_T} ,
\ee
where $\Delta y$ is a small rapidity interval centered at $y=0$ and $T_{\rm kin}$ is the {\em effective} temperature at the kinetic freeze-out which takes into account a sizable radial expansion of the fireball. One checks that the distribution (\ref{p-pT-y-dis}) is normalized to $N_p$.

The number of deuterons is found as
\ba
\nn
N_D\equiv \int d^3p \, \frac{dN_D}{d^3 p} = {\cal A}_2 \int d^3p \Big(\frac{dN_D}{d^3 p}\Big)^2
= \frac{2\, N_p^2}{\pi \, \Delta y} \, 
\frac{{\cal A}_2}{T_{\rm kin}(T_{\rm kin} + m)^2} .
\ea

To obtain the formation rate ${\cal A}_2$ in a simple analytic form, we do not use the Hulth\' en wave function of a deuteron, as we did in \cite{Mrowczynski:1992gc}, but we choose not only the Gaussian parameterizations of the source function but also of  $|\phi_D ({\bf r}) |^2$. Thus, we get 
\be
\label{A-Gauss}
{\cal A}_2 = \frac{3}{4} \frac{\pi^{3/2}}{(R^2_{\rm kin} + \frac{2}{3} R^2_D)^{3/2}} .
\ee 
where $R_D$ is the root-mean-square radius of a deuteron. In our original paper \cite{Mrowczynski:2016xqm} the factor $2/3$ in front of $R^2_D$ in Eq.~(\ref{A-Gauss}) was missing which influenced though insignificantly our numerical results. 

Using the formula (\ref{A-Gauss}), the ratio of the deuteron to proton yields equals
\ba
\label{d-p-coal-1}
\frac{D}{p} &=&
\frac{3 \sqrt{\pi} \, \lambda_p}{ (2\pi)^{3/2} \Delta y} \, 
\frac{V_{\rm chem} }{(R^2_{\rm kin} + \frac{2}{3} R^2_D)^{3/2}} \,
 \frac{(m T_{\rm chem})^{3/2}   }{T_{\rm kin}(T_{\rm kin} +m)^2} 
\\[2mm]  \nn
&&~~~~~~~~~~~~~~~~~~\times
e^{-\beta_{\rm chem}m} \,
\Bigg(1 + \frac{15 T_{\rm chem}}{8 m}  +{\cal O}\bigg( \frac{T^2_{\rm chem}}{m^2}\bigg) \Bigg),
\ea 
where the number of protons $N_p$ is assumed to be the same as in the thermal model and it is given by Eq.~(\ref{Np-chem-approx}). 

The ratio of the ratios (\ref{d-p-coal-1}) and (\ref{d-p-TM}), which is denoted as $Q$, equals the ratio of deuteron yields in the coalescence and thermal models because the proton yield is assumed to be the same in both approaches. The ratio $Q$ equals
\ba
\label{Q-def}
Q &\equiv& \frac{\big(D/p\big)_{\rm CM}}{\big(D/p\big)_{\rm TM}} = \frac{D_{\rm CM}}{D_{\rm TM}}
\\[2mm]  \nn
&=& \frac{\lambda_p^2}{2^{5/2} \pi \,\Delta y} \,
\frac{V_{\rm chem} }{(R^2_{\rm kin} + \frac{2}{3} R^2_D)^{3/2}}\,
\frac{(mT_{\rm chem})^{3/2}}{T_{\rm kin}(T_{\rm kin} + m)^2} 
\Bigg(1 + \frac{45 T_{\rm chem}}{16 m} +{\cal O}\bigg( \frac{T^2_{\rm chem}}{m^2}\bigg) \Bigg).
\ea
In the next section, after estimating the parameters which enter Eq.~(\ref{Q-def}), a magnitude of the ratio $Q$ is computed. 

\subsection{Discussion of $D/p$ ratio}
\label{sec-discussion}

\begin{table}[b]
\caption{\label{table-parameters} The ratio $Q$ and the centrality dependent parameters of Pb-Pb collisions at $\sqrt{s_{\rm N\! N}}=2.76$ TeV.  The numbers in the first three columns are taken from the experimental study \cite{Adam:2015vda}. The parameters $V_{\rm chem}$, $T_{\rm kin}$ and $R_{\rm kin}$ are estimated as explained in the text. The ratio  $Q$ is given by Eq.~(\ref{Q-def}). }
\center
\begin{tabular}{ c c c c c c c }
\hline 
\\
Centrality & $N_D$ & $\langle p_T \rangle$ & $V_{\rm chem}$ & $T_{\rm kin}$ & $R_{\rm kin}$ & $Q$
\\
 &  &[GeV] & $[{\rm fm}^3]$ & [MeV] & [fm]&  
\\[1mm]
\hline
\\[-2mm]
0 - 10\%   & 0.098 & 2.12 &    3 590 & 900 & 4.3 & 0.52
\\[1mm]
10 - 20\% & 0.076 & 2.07 &    2 780 & 890 & 3.7 & 0.61
\\[1mm]
20 - 40\% & 0.048 & 1.92 &    1 760 & 850 & 3.1 & 0.65
\\[1mm]
40 - 60\% & 0.019 & 1.63 &~~~696 & 760 & 2.5 & 0.51
\\
\hline
\end{tabular}
\end{table}

The $D/p$ ratio found within the thermal model (\ref{d-p-TM}) is determined by the proton mass $m$, the temperature of chemical freeze-out $T_{\rm chem}$ and the parameter $\lambda_p$ which we choose in such a way that at $T_{\rm chem} = 156$ MeV the ratio (\ref{d-p-TM}) reproduces the experimental value $3.6 \times 10^{-3}$ measured in Pb-Pb collisions at $\sqrt{s_{\rm N\! N}}=2.76$ TeV \cite{Adam:2015vda}. Thus, one finds $\lambda_p = 2.51$ which is used further on.

To obtain the $D/p$ ratio within the coalescence model (\ref{d-p-coal-1}), one needs, except $m$, $T_{\rm chem}$ and $\lambda_p$, the values of $\Delta y$, $R_D$,  $V_{\rm chem}$, $R_{\rm kin}$ and $T_{\rm kin}$.  The measurement \cite{Adam:2015vda} was performed in the rapidity window $\Delta y=1$. The root-mean-square radius of the deuteron is $R_D = 2$ fm \cite{Angeli:2013epw}. $V_{\rm chem}$ can be found from the deuteron analog of the formula (\ref{Np-chem-approx}), using the measured number of deuterons at different collision centralities given in \cite{Adam:2015vda}. 

In our original paper \cite{Mrowczynski:2016xqm} we used the femtoscopic pion data \cite{Adam:2015vna} to get a value of $R_{\rm kin}$. However, the experimental analysis \cite{Adam:2015vja} shows that the radii of pion sources are significantly bigger than those of proton sources. Therefore, we use here the radii of (anti-)proton sources inferred from proton-proton and antiproton-antiproton correlation functions at the smallest transverse momentum \cite{Adam:2015vja}. Since the data presented in \cite{Adam:2015vja} are for different centrality bins than those in \cite{Adam:2015vda} we have performed a linear interpolation or extrapolation to get values of $R_{\rm kin}$ for centrality bins given in the first column of the Table~\ref{table-parameters}.

The parameter $T_{\rm kin}$ from the formula (\ref{Q-def}) is the {\em effective} temperature at kinetic freeze-out which takes into account a sizable radial expansion of the fireball.  To determine $T_{\rm kin}$ we express it through the mean transverse momentum of deuterons $\langle p_T \rangle$ given in \cite{Adam:2015vda} using the formula
\ba
\label{pT}
\langle p_T \rangle &\equiv& \frac{\int_0^\infty dp_T p_T^2 \, e^{-\beta_{\rm kin}\sqrt{4m^2 + p_T^2}}}
{\int_0^\infty dp_T p_T \, e^{-\beta_{\rm kin}\sqrt{4m^2 + p_T^2}}} 
= \frac{4m^2}{T_{\rm kin}(1 + 2 \beta_{\rm kin}m)} \, e^{2\beta_{\rm kin}m} K_2(2\beta_{\rm kin}m).
\ea 
Since the effective kinetic temperature is comparable to the nucleon mass, the approximation (\ref{approx-K2}) cannot be applied to the Macdonald function in Eq.~(\ref{pT}). 

In Table~\ref{table-parameters} we list the values of the ratio $Q$ defined by Eq.~(\ref{Q-def}) for the four collision centralities together with the parameters of Pb-Pb collisions at $\sqrt{s_{\rm N\! N}}=2.76$ TeV. The predictions of the models are seen to differ by the factor smaller than 2 for all four centralities where number of deuterons grows five times from the peripheral to central collisions. 

We conclude that the two models indeed predict very similar yields of deuterons. We do not see any deeper reason for the similarity but it is also not accidental, this is a game of numbers -- the parameters, which characterize the produced matter at the chemical and kinetic freeze-out, are correlated with each other in a specific way.

\section{How to falsify one of the models?}

As we discussed in the previous section, the thermal and coalescence model predict similar yields of light nuclei. One asks whether one of the models can be falsified. For this purpose we need a situation that the models give quantitatively different predictions. Below we discuss two such situations and two proposals to distinguish the models. The first one is to measure the yield of exotic nuclide $^4{\rm Li}$ and compare it to that of $^4{\rm He}$. The ratio of yields of $^4{\rm Li}$ to $^4{\rm He}$ is different in the thermal and coalescence models  \cite{Mrowczynski:2016xqm,Bazak:2018hgl,Bazak:2020wjn}. The second proposal is to measure a hadron-deuteron correlation function which appears to carry information \cite{Mrowczynski:2019yrr} whether a deuteron is emitted from a fireball together with all other hadrons, as assumed in the thermal model, or a deuteron is formed only after nucleons are emitted, as in the coalescence model. Another version of the second proposal is to measure a hadron$-^3{\rm He}$ correlation function which can tell us about an origin of $^3{\rm He}$ \cite{Bazak:2020wjn}.

\subsection{$^4{\rm Li}$ vs. $^4{\rm He}$}
\label{sec-4Li-4He}

The mass of exotic nuclide $^4{\rm Li}$ is close to the mass of $^4{\rm He}$. However, there are five spin states of $^4{\rm Li}$, which has spin 2, and only one spin state of $^4{\rm He}$, which has zero spin. Consequently, the thermal model predicts about five times bigger yield of $^4{\rm Li}$ than that of $^4{\rm He}$.  If one takes into account the mass difference of $^4{\rm Li}$ and $^4{\rm He}$ the ratio of yields is reduced from 5 to 4.3 at the temperature of 156 MeV. Since the nuclide $^4{\rm Li}$ is loose while $^4{\rm He}$ is well bound and compact, the coalescence model is expected to predict a significantly smaller ratio of yields of $^4{\rm Li}$ to $^4{\rm He}$ than the thermal model. So, let us derive the ratio which in the coalescence model is given by the ratio of the formation rates to be computed according to the formula (\ref{A-form-rate}) with $A=4$. 

The modulus squared of the wave function of $^4{\rm He}$ is chosen as
\be
\label{alpha-wave-fun}
|\Psi_{\rm He}({\bf r}_1,{\bf r}_2, {\bf r}_3, {\bf r}_4) |^2 = C_\alpha 
e^{- \alpha ({\bf r}_{12}^2 + {\bf r}_{13}^2 + {\bf r}_{14}^2 + {\bf r}_{23}^2 + {\bf r}_{24}^2 + {\bf r}_{34}^2)},
\ee
where $C_\alpha$ is the normalization constant,  ${\bf r}_{ij} \equiv {\bf r}_i - {\bf r}_j$ and $\alpha$ is the parameter related to the RMS radius $R_\alpha$ of $^4{\rm He}$. Further calculations are performed using the Jacobi variables defined as
\be 
\label{Jacobi-4}
\left\{ \begin{array}{ll}
{\bf R} \equiv \frac{1}{4}({\bf r}_1+{\bf r}_2+{\bf r}_3 + {\bf r}_4) ,
\\[2mm]
{\bf x} \equiv {\bf r}_2-{\bf r}_1,
\\[2mm]
{\bf y} \equiv {\bf r}_3-\frac{1}{2}({\bf r}_1+{\bf r}_2) ,
\\[2mm]
{\bf z} \equiv {\bf r}_4 -\frac{1}{3}({\bf r}_1+{\bf r}_2+{\bf r}_3) ,
\end{array} \right.
~~~~~~~~~~~~~~~~~~
\left\{ \begin{array}{ll}
{\bf r}_1 = {\bf R} -\frac{1}{2} {\bf x} - \frac{1}{3} {\bf y} - \frac{1}{4} {\bf z} ,
\\[2mm]
{\bf r}_2 = {\bf R} + \frac{1}{2}{\bf x} - \frac{1}{3} {\bf y} - \frac{1}{4} {\bf z} ,
\\[2mm]
{\bf r}_3 = {\bf R} + \frac{2}{3} {\bf y}- \frac{1}{4} {\bf z} ,
\\[2mm]
{\bf r}_4 = {\bf R} + \frac{3}{4} {\bf z} ,
\end{array} \right.
\ee
which have the nice property that the sum of squares of particles' positions and the sum of squares of differences of the positions are expressed with no mixed terms of the Jacobi variables that is
\ba
\label{Jacobi-property-1}
{\bf r}_1^2 + {\bf r}_2^2 + {\bf r}_3^2 + {\bf r}_4^2 
= 4 {\bf R}^2 + \frac{1}{2} {\bf x}^2 + \frac{2}{3} {\bf y}^2 + \frac{3}{4} {\bf z}^2 ,
\\ \label{Jacobi-property-2}
{\bf r}_{12}^2 + {\bf r}_{13}^2 + {\bf r}_{14}^2 + {\bf r}_{23}^2 + {\bf r}_{24}^2 + {\bf r}_{34}^2 = 2 {\bf x}^2 + \frac{8}{3} {\bf y}^2 + 3 {\bf z}^2 .
\ea
With the help of relations (\ref{Jacobi-property-1}) and (\ref{Jacobi-property-2}), one easily finds 
\be
\label{C-allpha-alpha}
C_\alpha = \frac{2^6}{V} \Big(\frac{\alpha}{\pi}\Big)^{9/2}, 
~~~~~~~~~~ 
\alpha = \frac{3^2}{2^5 R_\alpha^2}  ,
\ee
where $V$ is the normalization volume of the plane wave describing a free motion of the center of mass.

Substituting the formulas (\ref{Gauss-source}) and (\ref{alpha-wave-fun}) with the parameters (\ref{C-allpha-alpha}) into Eq.~(\ref{A-form-rate}), one finds the coalescence rate of  $^4{\rm He}$ as
\be
\label{alpha-rate}
{\cal A}_4^{\rm He} =  \frac{\pi^{9/2}}{2^{9/2}}
\frac{1}{\big(R_s^2 +\frac{4}{9} R_\alpha^2\big)^{9/2}} ,
\ee 
where the spin and isospin factors have been included. Since $^4{\rm He}$ is the state of zero spin and zero isospin, the factors are 
$g_S = g_I = 1/2^3$ because there are $2^4$ spin and $2^4$ isospin states of four nucleons and there are two zero spin and two zero isospin states. 

The stable isotope $^6{\rm Li}$ is a mixture of two cluster configurations $ ^4{\rm He}\!-\! ^2{\rm H}$ and $ ^3{\rm He}\!-\! ^3{\rm H}$ \cite{Bergstrom:1979gpv}. Since $^4{\rm Li}$ decays into $p + ^3{\rm He}$, we assume that it has the cluster structure $p\!-\! ^3{\rm He}$. Following \cite{Bergstrom:1979gpv} we parametrize the modulus squared of the wave function of $^4{\rm Li}$, which is approximately treated here as a stable nucleus, as
\ba
\label{Li-wave-fun}
|\Psi_{\rm Li}({\bf r}_1,{\bf r}_2, {\bf r}_3, {\bf r}_4) |^2 = 
C_{\rm Li} \, e^{-\beta ({\bf r}_{12}^2 + {\bf r}_{13}^2 + {\bf r}_{23}^2 )} 
{\bf z}^4 e^{-\gamma {\bf z}^2} \, |Y_{lm}(\Omega_{\bf z})|^2,
\ea
where the nucleons number 1, 2 and 3 form the $^3{\rm He}$ cluster while the nucleon number 4 is the proton; ${\bf z}$ is the Jacobi  variable (\ref{Jacobi-4}); $Y_{lm}(\Omega_{\bf z})$ is the spherical harmonics related to the rotation of the vector ${\bf z}$ with quantum numbers $l,m$. The summation over $m$ is included in the spin factor $g_S$. 

Using the Jacobi variables, one computes the constant $C_{\rm Li}$ together with the parameters $\beta$ and $\gamma$ as
\be
C_{\rm Li} = \frac{2^4 3^{1/2} \beta^3 \gamma^{7/2}}{5 \pi^{7/2} V}, ~~~~~~~~~ 
\beta = \frac{1}{3 R_c^2} , ~~~~~~~~~ 
\gamma = \frac{21}{2^3(4R_{\rm Li}^2 - 3R_c^2)} ,
\ee
where $R_c$ and $R_{\rm Li}$ are the root-mean-square radii of the cluster $^3{\rm He}$ and of the nuclide $^4{\rm Li}$,  respectively.

Substituting the formulas (\ref{Gauss-source}) and (\ref{Li-wave-fun}) into Eq.~(\ref{A-form-rate}), one finds the coalescence rate of  $^4{\rm Li}$ as
\be
\label{Li-rate}
{\cal A}_4^{\rm Li} 
= \frac{15 \pi^{9/2}}{2^{13/2}}  
 \frac{R_s^4}{\big(R_s^2 +\frac{1}{2} R_c^2\big)^3 
\big(R_s^2 +\frac{4}{7} R_{\rm Li}^2 - \frac{3}{7} R_c^2\big)^{7/2}} ,
\ee 
where the isospin and spin factors are computed as follows. The nuclide has the isospin $I =1,~I_z = 1$ and thus the isospin factor is $g_I = 3/2^4$ because there are three isospin states $I =1,~I_z = 1$ of four nucleons. The spin of the ground state of $^4{\rm Li}$ is 2 which can be arranged with the orbital angular momentum $l=1$ and $l=2$. We assume here that the cluster $^3{\rm He}$ has spin $1/2$ as the free nuclide  $^3{\rm He}$. When the spins of $^3{\rm He}$ and $p$ are parallel or antiparallel, the orbital number is $l=1$ or $l=2$, respectively. However, the ground state of $^4{\rm Li}$ is of negative parity which suggests that $l=1$. We note that the parity of a two-particle system is $P= \eta_1 \eta_2 (-1)^l$ and the internal parities $\eta_1, \,\eta_2$ of ${^1{\rm H}}$ and ${^3{\rm He}}$ are both positive. When $l=1$, the total spin of $^3{\rm He}$ and $p$ has to be one and there are $3^2$ such spin states of four nucleons. Consequently, there are $3^2$  angular momentum states with 5 states corresponding to spin 2 of  $^4{\rm Li}$ and thus $g_S = 5/2^4$.

\begin{figure}[t]
\centering
\includegraphics[scale=0.28]{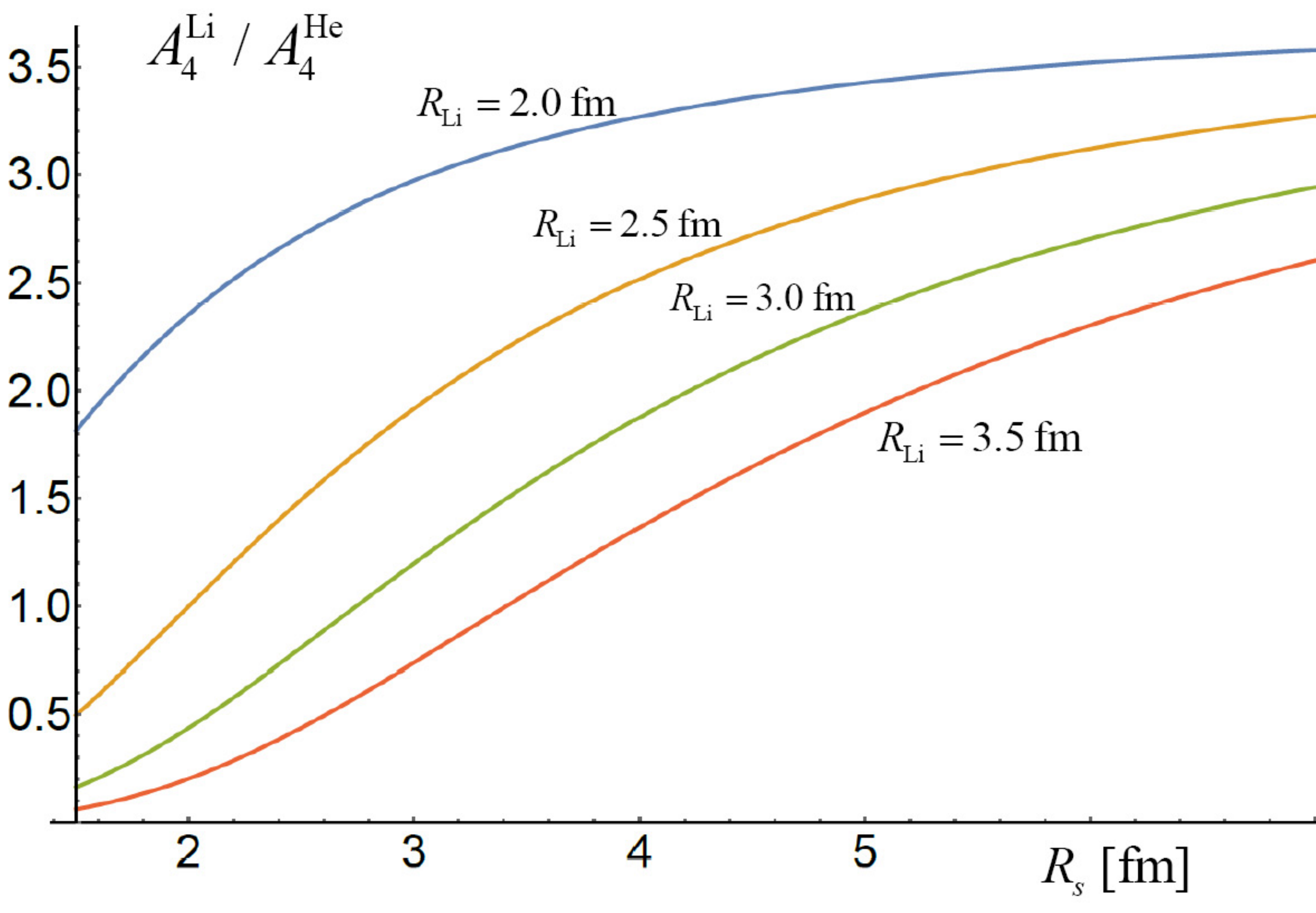}
\vspace{-1mm}
\caption{The ratio of formation rates of $^4{\rm Li}$ to $^4{\rm He}$ as a function of $R_s$ for four values of $R_{\rm Li} = 2.0,~2.5,~3.0$ and 3.5 fm.}
\label{Fig-ratio-rates-2}
\end{figure}

The ratio of the formation rates ${\cal A}_4^{\rm Li}$ and ${\cal A}_4^{\rm He}$ depends on four parameters: $R_s$, $R_\alpha$, $R_{\rm Li}$ and $R_c$. The fireball radius at the kinetic freeze-out $R_s$ can be inferred from the proton-proton correlation functions which have been precisely measured at LHC \cite{Adam:2015vja,Acharya:2018gyz}. The RMS radius of $^4{\rm He}$ is $R_\alpha = 1.68$ fm \cite{Angeli:2013epw} and the RMS radius of the cluster $^3{\rm He}$, which is identified with the radius of a free nucleus $^3{\rm He}$, is $R_c = 1.97$ fm \cite{Angeli:2013epw}. The radius $R_{\rm Li}$ is unknown but it is expected to be 2.5--3.5 fm. The ratio of ${\cal A}_4^{\rm Li}$ to ${\cal A}_4^{\rm He}$ is shown in Fig.~\ref{Fig-ratio-rates-2} as a function of $R_s$ for four values of $R_{\rm Li} = 2.0,~2.5,~3.0$ and 3.5 fm.

As already mentioned, the ratio of yields of $^4{\rm Li}$ to $^4{\rm He}$ in the thermal model equals about 5 and it is independent of the size of particle's source. Fig.~\ref{Fig-ratio-rates-2} shows that the ratio is significantly smaller in the coalescence model and it significantly depends on $R_s$  that is it depends on collision centrality. Therefore, performing measurements at several centralities it should be possible to quantitatively distinguish the coalescence mechanism of light nuclei production from the creation in a fireball. 

The nuclide $^4{\rm Li}$ is unstable and it decays into $p+{^3{\rm He}}$ with the width of 6 MeV \cite{NNDC}, see also \cite{Tilley:1992zz}. Therefore, the yield of $^4{\rm Li}$ can be experimentally obtained through a measurement of the $p\!-\!^3{\rm He}$ correlation function which is discussed in Sec.~\ref{sec-yield-Li}.

\subsection{Hadron-deuteron correlations}
\label{sec-h-D-correlations}

In this section we show that a hadron-deuteron correlation function carries information about the source of deuterons and allows one to determine whether a deuteron is directly emitted from the fireball or if it is formed afterwards. At first we derive the hadron-deuteron correlation function treating a deuteron as in the thermal model, that is as an elementary particle emitted from a source together with all other hadrons. Further on a deuteron is treated as a neutron-proton bound state formed at the same time when the hadron-deuteron correlation is generated. 

\subsubsection{Deuteron as an elementary particle}
\label{sec-deuteron-elementary}

The $h\!-\!D$ correlation function $\mathcal{R}$ is defined as
\be
\frac{dN_{hD}}{d^3p_h  d^3p_D} = \mathcal{R}({\bf q}) \, \frac{dN_h}{d^3p_h}  \frac{dN_D}{d^3p_D},
\ee
where $\frac{dN_h}{d^3p_h}$, $\frac{dN_D}{d^3p_D}$ and $\frac{dN_{hD}}{d^3p_h d^3p_D}$ are number densities of $h$, $D$ and $h\!-\!D$ pairs with momenta ${\bf p}_h$, ${\bf p}_D$ and $({\bf p}_h, {\bf p}_D)$; ${\bf q}$ is the relative momentum of $h$ and $D$ in their center-of-mass frame. If the correlation results from quantum statistics and/or final-state interactions, the correlation function is known to be \cite{Koonin:1977fh,Lednicky:1981su}
\be
\label{def-fun-cor-pD}
\mathcal{R}({\bf q}) = \int d^3 r_h \, d^3 r_D \, 
D({\bf r}_h) \, D({\bf r}_D) |\psi_{\bf q} ({\bf r}_h,{\bf r}_D)|^2 ,
\ee
where the source function $D({\bf r})$ is, as previously, the probability distribution of emission points and $\psi_{\bf q}({\bf r}_h,{\bf r}_D)$ is the wave function of the hadron and deuteron in a scattering state. 

Let us eliminate the center-of-mass motion of the $h\!-\!D$ pair in a non-relativistic manner. We introduce the center-of-mass variables 
\be
\label{CM-2-variables}
{\bf R} \equiv \frac{m_h{\bf r}_h + m_D{\bf r}_D }{M} ,
~~~~~~
{\bf r} \equiv  {\bf r}_h - {\bf r}_D ,
\ee
where $M \equiv m_h + m_D$, and we write down the wave function as $\psi_{\bf q}({\bf r}_h,{\bf r}_D) = e^{i {\bf R} {\bf P}}\phi_{\bf q} ({\bf r})$ with ${\bf P}$ being the momentum of the center of mass. The correlation function (\ref{def-fun-cor-pD}) is then found to be
\be
\label{fun-corr-relative}
\mathcal{R}({\bf q}) = \int d^3 r \, D_r ({\bf r}) |\phi_{\bf q}({\bf r})|^2 ,
\ee
where the `relative' source is 
\be
\label{D-r-def}
D_r({\bf r}) \equiv \int d^3 R \, D \Big ({\bf R} + \frac{m_D}{M}{\bf r}\Big) 
\, D \Big ({\bf R}-\frac{m_h}{M} {\bf r}\Big) .
\ee
With the Gaussian single-particle source function (\ref{Gauss-source}), the relative source function equals
\be
\label{D-r-Gauss}
D_r({\bf r})= \Big({\frac{1}{4 \pi R_s^2}}\Big)^{3/2} e^{ -\frac{{\bf r}^2}{4R_s^2} } ,
\ee
which is independent of particle masses even so the variable ${\bf R}$ given by Eq.~(\ref{CM-2-variables}) depends on $m_h$ and $m_D$. 

The single particle source function (\ref{Gauss-source}) is assumed to be independent of particle's momentum and particle's mass. This is not quite right as, in general, a source radius depends on both particle's mass $m$ and momentum. More precisely, it scales with the particle's transverse mass $m_\perp \equiv \sqrt{m^2 + p_\perp^2}$.  For the case of one-dimensional analysis relevant for our study, the effect is well seen in Fig.~8. of Ref.~\cite{Adam:2015vja} where experimental data on Pb-Pb collisions at LHC, which are of particular interest for us, are presented. The dependence of the source radius on $m_\perp$ is evident when we deal with pions and $m_\perp \lesssim 0.9~{\rm GeV}$. However, the dependence becomes much weaker for protons when $m_\perp \gtrsim 1.0~{\rm GeV}$. The figure shows that the radius of proton source tends to decrease in central Pb-Pb  collisions when $m_\perp $ grows from 1.1 GeV to 1.7 GeV but the decrease is not seen for the collision centrality $10-30\%$ nor $30-50\%$. The behavior is well understood as the decrease of the source radius with growing $m_\perp $ is caused by the collective radial flow which is stronger in central than in peripheral collisions. 

In case of proton-deuteron correlations, which are of our particular interest, the interval of $m_\perp $ from 1 to 2 GeV is of crucial importance. The experimental data from non-central collisions, which are presented in Fig.~8  of Ref.~\cite{Adam:2015vja}, show no dependence of the source radius on $m_\perp $ in the interval. Since we are interested in rather peripheral collisions, where the source radii are sufficiently small and the effect we suggest to measure is significant, it is legitimate to assume that the source radius is independent of particle's transverse mass. 

If the Coulomb interaction is absent but there is a short-range strong interaction, the wave function can be chosen, as proposed in \cite{Lednicky:1981su}, in the asymptotic scattering form
\be 
\label{scatt-wave-fun}
\phi_{\bf q}({\bf r}) = e^{iqz}+f(q)\frac{e^{iqr}}{r} ,
\ee
where $q \equiv |{\bf q}|$ and $f(q)$ is the $s-$wave (isotropic) scattering amplitude. 

With the source function (\ref{D-r-Gauss}) and the wave function (\ref{scatt-wave-fun}), the correlation function (\ref{fun-corr-relative}) equals
\ba
\nn
\label{fun-corr-final}
\mathcal{R}(q) 
&=& 1 + \frac{1}{2R_s^2}|f(q)|^2 - \frac{1 -  e^{-4R_s^2 q^2}}{2R_s^2 q} \, \Im f(q) 
\\[2mm]
&& ~~~~~~~~~~~~~~~~~~~~+ \;
\frac{1}{2 \pi^{1/2}R_s^3 q} \, \Re f(q)  \int^\infty_0 dr \,e^{-\frac{r^2}{4R_s^2}} \sin(2qr) .
\ea
The remaining integral needs to be taken numerically.  The formula (\ref{fun-corr-final}) has been repeatedly used to compute correlation functions of various two-particle systems. 

When one deals with charged particles, the formula (\ref{scatt-wave-fun}) needs to be modified because the long-range electrostatic interaction influences both the incoming and outgoing waves. However, the Coulomb effect can be approximately taken into account \cite{Gmitro:1986ay} by multiplying the correlation function by the Gamow factor that equals
\be 
\label{Gamow}
G(q) = \pm {2 \pi \over a_B q} \,
{1 \over {\rm exp}\big(\pm {2 \pi \over a_B q}\big) - 1} ,
\ee
where the sign $+$ ($-$) is for the repelling (attracting) particles and $a_B$ is the Bohr radius of the pair.

\subsubsection{Deuteron as a bound state}
\label{sec-deuteron-bound}

Let us now derive the $h\!-\!D$ correlation function treating the deuteron as a neutron-proton bound state created due to final-state interactions similarly to the $h\!-\!D$ correlation. Then, the correlation function is defined as
\be
\frac{dN_{hD}}{d^3p_h \, d^3p_D} = \mathcal{R}({\bf q}) \, \mathcal{A}_2 \,
\frac{dN_h}{d^3p_h} \frac{dN_n}{d^3p_n} \frac{dN_p}{d^3p_p},
\ee
where ${\bf p}_n = {\bf p}_p = {\bf p}_D/2$ and ${\bf q}$ is the relative momentum of $h$ and $D$. The deuteron formation rate $\mathcal{A}_2$ is given by the formula (\ref{A-form-rate}) with $A=2$. The correlation function multiplied by the deuteron formation rate equals
\be
\label{fun-corr-pi-D-A}
\mathcal{R}({\bf q}) \, \mathcal{A}_2 
= \frac{3}{4} (2 \pi )^3 \int d^3 r_h \, d^3 r_n \, d^3 r_p \, D({\bf r}_n) \, D({\bf r}_p) 
D({\bf r}_h) 
|\psi_{h n p}({\bf r}_h, {\bf r}_n, {\bf r}_p)|^2 ,
\ee
where $\psi_{h n p}({\bf r}_h, {\bf r}_n, {\bf r}_p)$ is the wave function of a $h\!-\!D$ system. The spin factor $3/4$ has the same origin as that in Eq.~(\ref{d-form-rate}). 

To compute the integral in Eq.~(\ref{fun-corr-pi-D-A}), we introduce the Jacobi variables of a three-particle system 
\be 
\label{Jacobi-3}
\left\{ \begin{array}{ll}
{\bf R}\equiv \frac{m_n{\bf r}_n + m_p{\bf r}_p + m_h {\bf r }_h}{M} ,
\\[1mm]
{\bf r}_{np} \equiv  {\bf r}_n-{\bf r}_p ,
\\[1mm]
{\bf r}_{h D} \equiv  {\bf r}_h - \frac{m_n {\bf r}_n + m_p {\bf r}_p}{m_D} ,
\end{array} \right.
\ee
with $M \equiv m_n + m_p +m_h$, $m_D \equiv m_n + m_p$ and we write down the wave function as
\be
\label{wave-fun-hnp}
\psi_{h n p}({\bf r}_h, {\bf r}_n, {\bf r}_p) = e^{i{\bf P}{\bf R}} \, \psi_{h D}^{\bf q}({\bf r}_{h D}) \, 
\varphi_D ({\bf r}_{np}) ,
\ee
where $\psi_{h D}^{\bf q}({\bf r}_{h D})$ and $\varphi_D ({\bf r}_{np})$ are the wave functions of the relative motion of $p$ and $D$ and of internal motion of $D$, respectively. 

Using the Gaussian source (\ref{Gauss-source}), the integral over the center-of-mass position ${\bf R}$  in Eq.~(\ref {fun-corr-pi-D-A}) gives
\be
\label{relative-sources-3}
\int d^3 R \,D ({\bf r}_n) \, D({\bf r}_p) \, D({\bf r}_h)
=   D_r({\bf r}_{np})  \, D_{3r}({\bf r}_{h D}) ,
\ee
where $ D_r({\bf r})$ is again given by Eq.~(\ref{D-r-Gauss}) and the normalized function $ D_{3r}({\bf r})$ equals
\be
\label{source-r-pi-D}
\mathcal{D}_{3r}({\bf r}) = \Big(\frac{1}{3 \pi R_s^2} \Big)^{3/2} e^{-\frac{{\bf r}^2}{3R_s^2}} .
\ee

As a result of the integration over ${\bf R}$ in the right-hand-side of Eq.~(\ref{fun-corr-pi-D-A}), the formation rate, which is given by Eq.~(\ref{d-form-rate}) factors out. Consequently, the rate, which is also present in the left-hand-side of Eq.~(\ref{fun-corr-pi-D-A}), drops out and the correlation function equals
\be
\label{fun-corr-pi-D-bound}
\mathcal{R}({\bf q}) = \int d^3 r_{h D} \, 
D_{3r}({\bf r}_{h D}) \, |\psi_{h D}^{\bf q} ({\bf r}_{h D})|^2.
\ee

The formula (\ref{fun-corr-pi-D-bound}) has the same form as (\ref{fun-corr-relative}) but the source function differs. When deuterons are directly emitted from the fireball as `elementary' particles the radius of deuteron source is the same as the radius of proton source. When deuterons are formed only after emission of nucleons from the fireball, the source becomes bigger because the  deuteron formation is a process of spatial extent. More quantitatively, the source radius of deuterons treated as bound states is bigger by the factor $\sqrt{4/3}\approx 1.15$ than that of `elementary' deuterons. 

\subsubsection{$p\!-\!D$ correlation function}

We have first considered the $K^-\!\!-\!D$ correlation function which has appeared not very sensitive to a source radius because the strong interaction of $K^-$ and $D$ is rather weak that is the scattering length is short. The $p\!-\!D$ system is better suited for our purpose. 

Since the $p\!-\!D$ pair can have spin 1/2 or 3/2 there are two interaction channels. The $s-$wave scattering lengths of $p\!-\!D$ scattering in the 1/2 and 3/2 channels are, respectively, 4.0 fm and 11.0 fm \cite{Black:1999duc}. As nucleons are assumed unpolarized, the $p\!-\!D$ correlation function is computed as the average  
\be
\label{ave-corr-p-D}
\mathcal{R}({\bf q}) = \frac{1}{3}\, \mathcal{R}^{1/2}({\bf q}) + \frac{2}{3}\, \mathcal{R}^{3/2}({\bf q}) ,
\ee
where the weights factors 1/3 and 2/3 reflect the numbers of spin states in the two channels. 

\begin{figure}[t]
\centering
\includegraphics[scale=0.29]{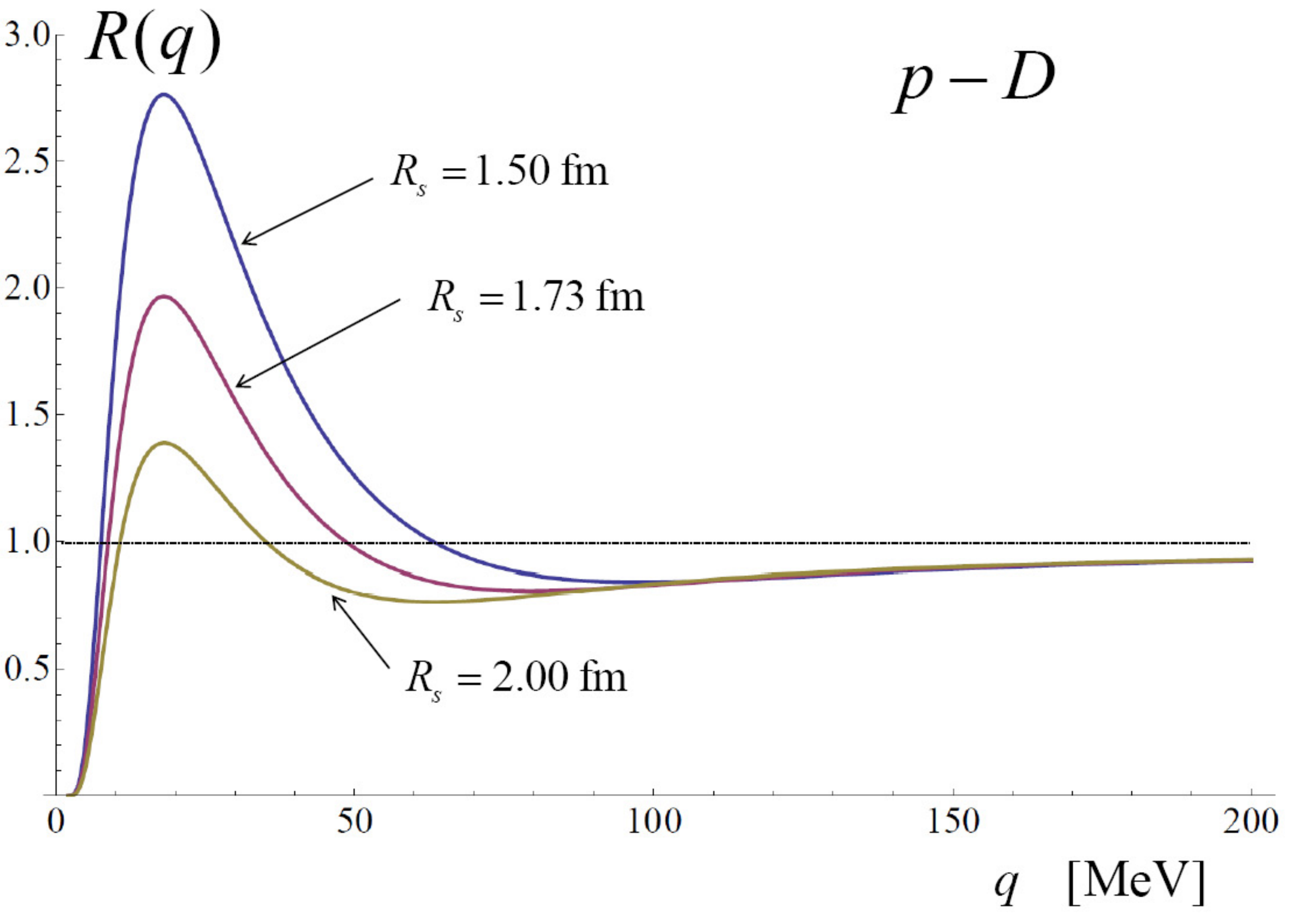}
\vspace{-2mm}
\caption{$p\!-\!D$ correlation function}
\label{fig-p-D}
\end{figure}

The average $p\!-\!D$ correlation function is shown in Fig.~\ref{fig-p-D} for three values of the source radius which are chosen in such a way that $R_s = 2.00 \; {\rm fm} = \sqrt{4/3}\cdot 1.73 \; {\rm fm} = \frac{4}{3}\cdot 1.50 \; {\rm fm}$. The function strongly depends on $R_s$. Therefore, it should be possible to infer the source radius from experimentally measured $p\!-\!D$ correlation function and compare it to $R_s$ obtained from the $p\!-\!p$ correlation function. If deuterons are directly emitted from the fireball, the radii of proton and deuteron sources are the same. If deuterons are formed due to final-state interactions, the radius of deuteron source is bigger by the factor $\sqrt{4/3}$. To distinguish the two scenarios the $p\!-\!D$ correlation function should be measured with such an accuracy that one can distinguish the two neighboring curves in  Fig.~\ref{fig-p-D}.

The dependence of the $p\!-\!D$ correlation function on $R_s$  becomes weaker as $R_s$ grows. Consequently, the analysis of higher $p_T$ particles from non-central events, when the sources are relatively small, is preferred. 

Our proposal to distinguish the scenario of deuterons directly emitted from the fireball from that of deuterons formed due to final-state interactions does not relay on an absolute value of the source size inferred from the $p\!-\!D$ correlation function but on a comparison of source size parameters inferred from the $p\!-\!D$ and $p\!-\!p$ correlation functions. Therefore, systematic uncertainties of the femtoscopic method, both experimental and theoretical, are not of crucial importance here, as they are expected to influence in a similar way the source parameters inferred from the $p\!-\!D$ and $p\!-\!p$ correlation functions. 

We note that the size of the proton source in $p\!-\!p$ collisions at LHC was measured with an experimental accuracy of 7\% where the statistical error is only 2\% \cite{Acharya:2018gyz}. Our proposal requires an accuracy better than 15\% which, however, does not include systematic experimental and theoretical uncertainties. Therefore, the required accuracy of the measurement seems achievable. 

It should be also noted that the $p\!-\!D$ correlation function was measured in heavy-ion collisions at low and intermediate collision energies that is from a few tens to a few hundreds of MeV/nucleon, see the review \cite{Ardouin:1996nc} and references therein. The $p\!-\!D$ correlations in Pb-Pb collisions at LHC are currently under study by the ALICE Collaboration \cite{Laura}.

\subsection{$p\!-\!^3{\rm He}$ correlations}
\label{sec-p-3He-correlation}

The $p\!-\!^3{\rm He}$ correlation function is interesting for two different reasons. It is needed to obtain the yield of $^4{\rm Li}$ but at the same time the correlation function carries information about a source of $^3{\rm He}$. It allows one to determine whether $^3{\rm He}$ is directly emitted from the fireball, as in the thermal model, or it is formed afterwards, as in the coalescence approach. 

\subsubsection{General formula of $p\!-\!^3{\rm He}$ correlation function}
\label{sec-gen-corr-fun}

If the nucleus $^3{\rm He}$ is treated as an elementary particle emitted from a source, the $p\!-\!^3{\rm He}$ correlation function is defined as
\be
\label{fun-corr-p-3He-elem}
\frac{dN_{p{^3{\rm He}}}}{d^3p_p  d^3p_{^3{\rm He}}} = \mathcal{R}({\bf q}) \, 
\frac{dN_p}{d^3p_p}  \frac{dN_{^3{\rm He}}}{d^3p_{^3{\rm He}}} ,
\ee
and it is given by Eq.~(\ref{fun-corr-relative}) where the source function (\ref{D-r-Gauss}) enters. The wave function $\phi_{\bf q}({\bf r})$, however, describes not the relative motion of $h$ and $D$ but of $p$ and $^3{\rm He}$.

Taking into account that the nucleus $^3{\rm He}$ is a bound state of $(p,p,n)$ formed due to final-state interactions at the same time when the correlation among ${^3{\rm He}}$ and $p$ is generated, the correlation function is defined as
\be
\label{N-p-3He}
\frac{d N_{p \,{^3{\rm He}}}}{d^3 p_p d^3 p_{^3{\rm He}}} 
= \mathcal{R}({\bf q})\,\mathcal{A}_3\,
\frac{d N_p}{d^3 p_p}
\frac{d N_N}{d^3(p_{^3{\rm He}}/3)}
\frac{d N_N}{d^3(p_{^3{\rm He}}/3)}
\frac{d N_N}{d^3(p_{^3{\rm He}}/3)},
\ee
where $\mathcal{A}_3$ is the formation rate of a nucleus $^3{\rm He}$ given by the formula (\ref{A-form-rate}) with $A=3$. The product of the formation rate and correlation function is
\ba
\nn
\mathcal{R}({\bf p}_{\rm p}, {\bf p}_{^3{\rm He}})\mathcal{A}_3 &=&
g_S g_I (2\pi)^6 \int d^3 {\bf r}_{\rm p} d^3 {\bf r}_{1}d^3 {\bf r}_{2}d^3 {\bf r}_{3} \, 
D({\bf r}_{\rm p})D({\bf r}_1)D({\bf r}_2)D({\bf r}_3)
\\[2mm] \label{RA}
&& ~~~~~~~~~~~~~~~~~~~~~~~~~~~~~~~~~~~~~~~~
\times |\psi_{{\rm p}\, ^3{\rm He}}({\bf r}_{\rm p},{\bf r}_1,{\bf r}_2, {\bf r}_3)|^2,
\ea
where $D({\bf r}_i)$ with $i= p,1, 2, 3$ is again the source function while $\psi_{{\rm p}\, ^3{\rm He}}({\bf r}_{\rm p},{\bf r}_1,{\bf r}_2, {\bf r}_3)$ is the wave function of  $p$ and ${^3{\rm He}}$. 

Let us compute $\mathcal{A}_3$. Using the Jacobi variables for a system of three particles with equal masses 
\be 
\label{Jacobi-3}
\left\{ \begin{array}{ll}
{\bf R} \equiv \frac{1}{3}({\bf r}_1+{\bf r}_2+{\bf r}_3) ,
\\[2mm]
{\bf x} \equiv {\bf r}_2-{\bf r}_1,
\\[2mm]
{\bf y} \equiv {\bf r}_3-\frac{1}{2}({\bf r}_1+{\bf r}_2) ,
\end{array} \right.
~~~~~~~~~~~~~~~~~~
\left\{ \begin{array}{ll}
{\bf r}_1 = {\bf R} -\frac{1}{2} {\bf x} - \frac{1}{3} {\bf y} ,
\\[2mm]
{\bf r}_2 = {\bf R} + \frac{1}{2}{\bf x} - \frac{1}{3} {\bf y} ,
\\[2mm]
{\bf r}_3 = {\bf R} + \frac{2}{3} {\bf y} ,
\end{array} \right.
\ee
and writing down the wave function of $^3{\rm He}$ as 
\be
\psi_{^3{\rm He}}({\bf r}_1, {\bf r}_2, {\bf r}_3) = e^{i{\bf P}{\bf R}} \, \phi_{^3{\rm He}}({\bf x}, {\bf y}),
\ee
with $\phi_{^3{\rm He}}({\bf x}, {\bf y})$ being the wave function of relative motion, the formation rate equals
\be
\label{A3}
\mathcal{A}_3 = g_S g_I (2\pi)^6 \int d^3 {\bf x}\,  d^3{\bf y}\,
 D_r({\bf x},{\bf y}) \, |\phi_{^3{\rm He}}({\bf x}, {\bf y})|^2,
\ee
where the normalized two-particle relative source function $D_r({\bf x},{\bf y})$ is
\ba
\nn
D_r({\bf x},{\bf y}) \equiv \int d^3R \, 
D\Big({\bf R} -\frac{1}{2} {\bf x} - \frac{1}{3} {\bf y}\Big) \,
D\Big({\bf R} +\frac{1}{2} {\bf x} - \frac{1}{3} {\bf y}\Big) \,
D\Big({\bf R} + \frac{2}{3} {\bf y}\Big) 
\\[2mm] \label{Dr-x-y}
= \frac{1}{\big(2\sqrt{3}\pi R_s^2\big)^3}\, 
e^{-\frac{{\bf x}^2}{4R_s^2} -\frac{{\bf y}^2}{3R_s^2}} .
\ea
The latter equality holds for the Gaussian parametrization (\ref{Gauss-source}). 

To derive the  $p\!-\!^3{\rm He}$ correlation function we use the Jacobi variables for a system of four particles defined by Eqs.~(\ref{Jacobi-4}) and we write down the wave function as 
\be
\label{wave-function-p-3He}
\psi_{{\rm p}\, ^3{\rm He}}({\bf r}_{\rm p},{\bf r}_1,{\bf r}_2, {\bf r}_3) 
= e^{i{\bf P}{\bf R}} \, \varphi_{\bf q} ({\bf z}) \, \phi_{^3{\rm He}}({\bf x}, {\bf y}),
\ee
where $\varphi_{\bf q} ({\bf z})$ is the wave function of relative motion in the center of mass of  $p\!-\!^3{\rm He}$ system. Computing the integral (\ref{RA}), one finds that $\mathcal{A}_3$ factors out and the correlation function equals
\be
\label{fun-cor-bound}
\mathcal{R}({\bf q})=\int d^3 {\bf r}\, D_{4r}({\bf r}) \,|\varphi_{\bf q}({\bf r})|^2 .
\ee
The source function $D_{4r}({\bf r})$ is defined through the equality
\ba
\nn
D_r({\bf x},{\bf y}) \, D_{4r}({\bf z})
&=& (2\pi)^6 \int d^3 {\bf R} \, 
D\Big({\bf R} -\frac{1}{2} {\bf x} - \frac{1}{3} {\bf y} - \frac{1}{4} {\bf z}\Big) 
\\[2mm] \label{Dr-4}
&\times&
D\Big({\bf R} +\frac{1}{2} {\bf x} - \frac{1}{3} {\bf y}- \frac{1}{4} {\bf z}\Big) \,
D\Big({\bf R} + \frac{2}{3} {\bf y}- \frac{1}{4} {\bf z}\Big) \,
D\Big({\bf R} + \frac{3}{4} {\bf z}\Big) ,
\ea
where the function $D_r({\bf x},{\bf y})$ is given by Eq.~(\ref{Dr-x-y}). Using the Gaussian parametrization (\ref{Gauss-source}), the normalized source function $D_{4r}({\bf r})$ is found as 
\be
\label{Dz}
 D_{4r}({\bf z}) = \Big(\frac{3}{8\pi R_s^2}\Big)^{3/2} e^{-\frac{3 {\bf z}^2}{8R_s^2}} .
\ee

The correlation functions (\ref{fun-corr-relative}) and (\ref{fun-cor-bound}) differ only due to the different source functions (\ref{D-r-Gauss}) and (\ref{Dz}), respectively. The source radius of nuclei $^3{\rm He}$ treated as bound states is bigger by the factor $\sqrt{3/2}\approx 1.22$ than that of  the `elementary' nuclides $^3{\rm He}$. When the source radius inferred from the $p\!-\!p$ correlation function is the same as the radius obtained from the $p\!-\!^3{\rm He}$ correlation function, it means that the nuclides $^3{\rm He}$ are directly emitted from the fireball. If the radius is bigger by $\sqrt{3/2}$, the nuclides are formed due to final-state interactions. The question is, however, whether the $p\!-\!^3{\rm He}$ correlation function is sensitive enough to the change of source radius from $R_s$ to $\sqrt{3/2}\,R_s$. The question is discussed in the next section.

\subsubsection{$p\!-\!^3{\rm He}$ correlation function}

The effect of $s-$wave scattering and Coulomb repulsion of $p$ and $^3{\rm He}$ is computed exactly as in the case of $p\!-\!D$ correlation function discussed in Sec.~\ref{sec-deuteron-elementary}. The difference is that both proton and $^3{\rm He}$ have spin 1/2 and there are singlet (spin zero) and triplet (spin one) channels of the $p\!-\!^3{\rm He}$ scattering. Consequently, the spin average $p\!-\!^3{\rm He}$ correlation function is given not by Eq.~(\ref{ave-corr-p-D}) but it equals 
\be
\mathcal{R}({\bf q})=\frac{1}{4}\mathcal{R}^s({\bf q})+\frac{3}{4}\mathcal{R}^t({\bf q}) .
\ee
The singlet and triplet scattering lengths are sizable \cite{Daniels:2010af} and are 
\be
a_s=11.1\; {\rm fm}, 
~~~~~~~~~~~~~~
a_t=9.05\;{\rm fm}.
\ee

\begin{figure}[t]
\centering
\includegraphics[scale=0.27]{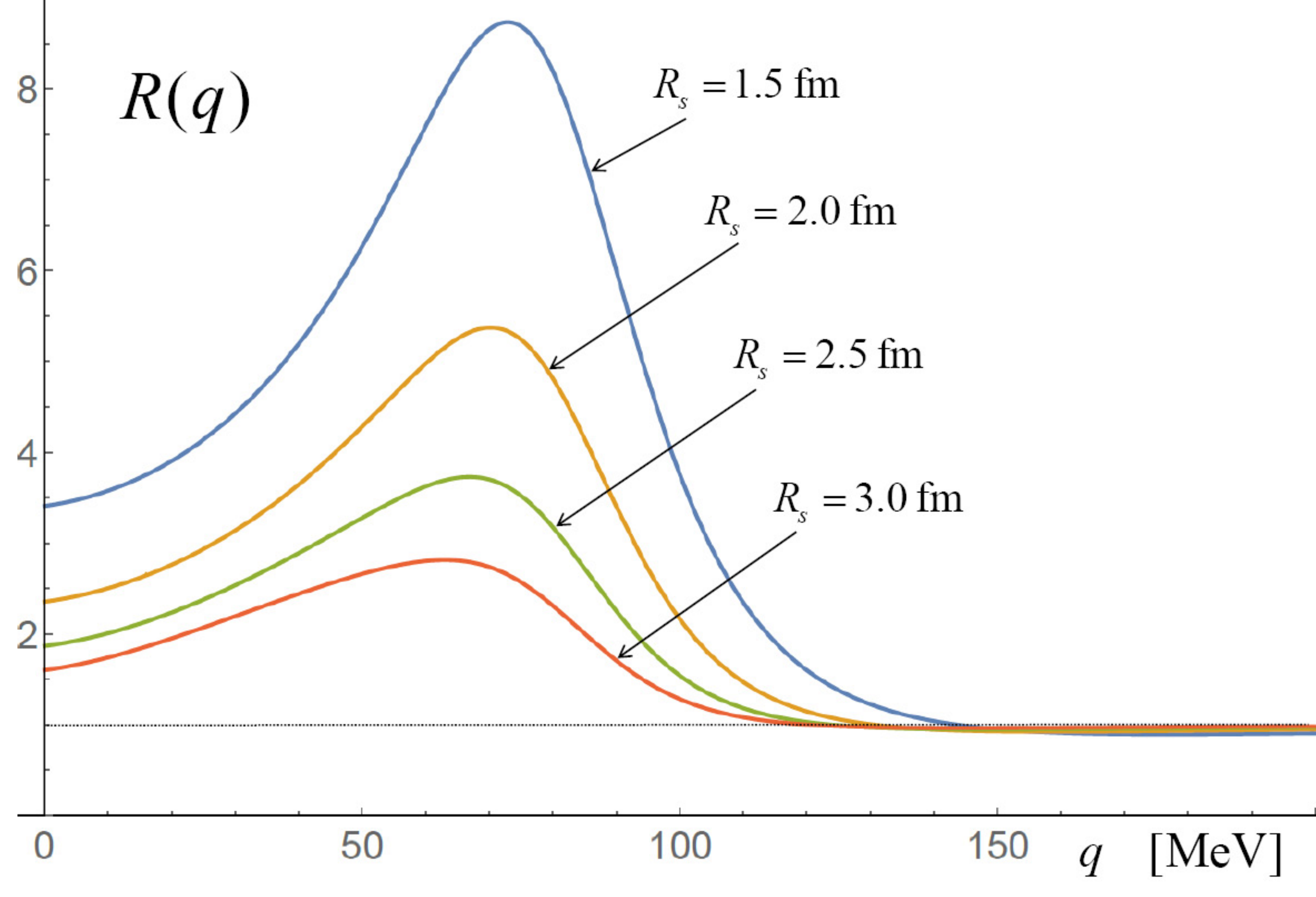}
\vspace{-2mm}
\caption{The $p\!-\!^3{\rm He}$ correlation function which takes into account only the resonance of $^4{\rm Li}$.}
\label{Fig-resonance-1}
\end{figure}

Our main interest is the of $p\!-\!^3{\rm He}$ resonance interaction due to the transient state of $^4{\rm Li}$. The resonance mass equals $\Delta_E = 4,07$ MeV above the sum of masses of proton and ${^3{\rm He}}$ and its width is $\Gamma = 6,03$ MeV \cite{Tilley:1992zz}. The amplitude, which takes into account the resonance scattering, is \cite{Landau-Lifshitz-1988}
\be
f(q,\theta)=f(q) + f_l^r(q) \, P_l(\cos\theta),
\ee
where $f(q)$ is the $s-$wave scattering amplitude,  and $f_l^r(q)$ is the resonance contribution with $l$ being the orbital momentum of the resonance, $P_l(\cos\theta)$ is the Legendre polynomial which for $l=1$ equals $P_1(\cos\theta) = \cos\theta$. As discussed in Sec.~\ref{sec-4Li-4He}, the orbital angular momentum of $^4{\rm Li}$ is $l=1$. 

The resonance amplitude is of the Breit-Wigner form
\be
\label{B-W-amplitude}
f_l^r(q) \equiv - \lambda_R\ \frac{2l+1}{q_0}\frac{\frac{1}{2}\Gamma}{E-E_0+\frac{1}{2}i\Gamma},
\ee
where $E_0$ and $\Gamma$ are the resonance energy and its width and the parameter $\lambda_R$, which is assumed to be real, controls a strength of the resonance. In case of the  $^4{\rm Li}$ resonance, the energy difference, which enters the amplitude, is 
\be
E-E_0 = \frac{q^2}{2\mu} -  \Delta_E ,
\ee
and the momentum $q_0$, which corresponds to the resonance peak, is $q_0=75.7$~MeV. 

When compared to the original formula (134.12) from the textbook \cite{Landau-Lifshitz-1988}, we have introduced the parameter $\lambda_R$ and have replaced the factor $(2l+1)/q$ by $(2l+1)/q_0$, where $q_0$ corresponds to the energy $E_0$, to avoid the divergence of the amplitude at $q=0$. The modification is legitimate as, strictly speaking, the amplitude is valid only in the vicinity of the resonance.  In our numerical calculations we assume that $\lambda_R=1$ but, as we discuss further on, the parameter $\lambda_R$ can be and should be inferred from experimental data. 

The correlation function, which is computed with the source function  (\ref{Dz}) and takes into account the resonance interaction, is
\ba
\nn
\mathcal{R}({\bf q}) 
&=& \mathcal{R}_0({\bf q}) + \Big(\frac{3}{8\pi R_s^2}\Big)^{3/2} |f_l^r(q)|^2 J_l 
\\[2mm] \label{fun-corr-gen-final}
&&~~~~~~~~+ \;  2 \Big(\frac{3}{8\pi R_s^2}\Big)^{3/2}
\Big(\Re f_l^r(q) \, \Re K_l(q) -  \Im f_l^r(q)  \, \Im K_l(q) \Big) ,
\ea
where $\mathcal{R}_0({\bf q})$ is the correlation function due to $s-$wave scattering only. For $l=1$ the coefficient $J_l$ and the function $K_l(q)$ are
\be
J_1 = \frac{2^{5/2} \pi^{3/2}}{3^{3/2}} \, R_s, 
\ee
\ba
\Re K_1(q) &=& - \frac{2\pi}{q} \int_0^\infty dr \, e^{-\frac{3{\bf r}^2}{8R_s^2}} \, \sin(2qr) 
+ \frac{4\pi}{q^2} \int_0^\infty dr \, e^{-\frac{3{\bf r}^2}{8R_s^2}} \;  
 \frac{\sin^2(qr)}{r}  ,
\\[2mm] 
\label{ImK1}
\Im K_1(q) &=& \frac{2^{3/2} \pi^{3/2} R_s}{3^{1/2}q} \Big( 1 + e^{- \frac{8 q^2 R_s^2}{3}} \Big) 
- \frac{2\pi}{q^2} \int_0^\infty dr \, e^{-\frac{3{\bf r}^2}{8R_s^2}} \;  
\frac{\sin(2qr)}{r} .
\ea

\begin{figure}[t]
\centering
\vspace{3mm}
\includegraphics[scale=0.27]{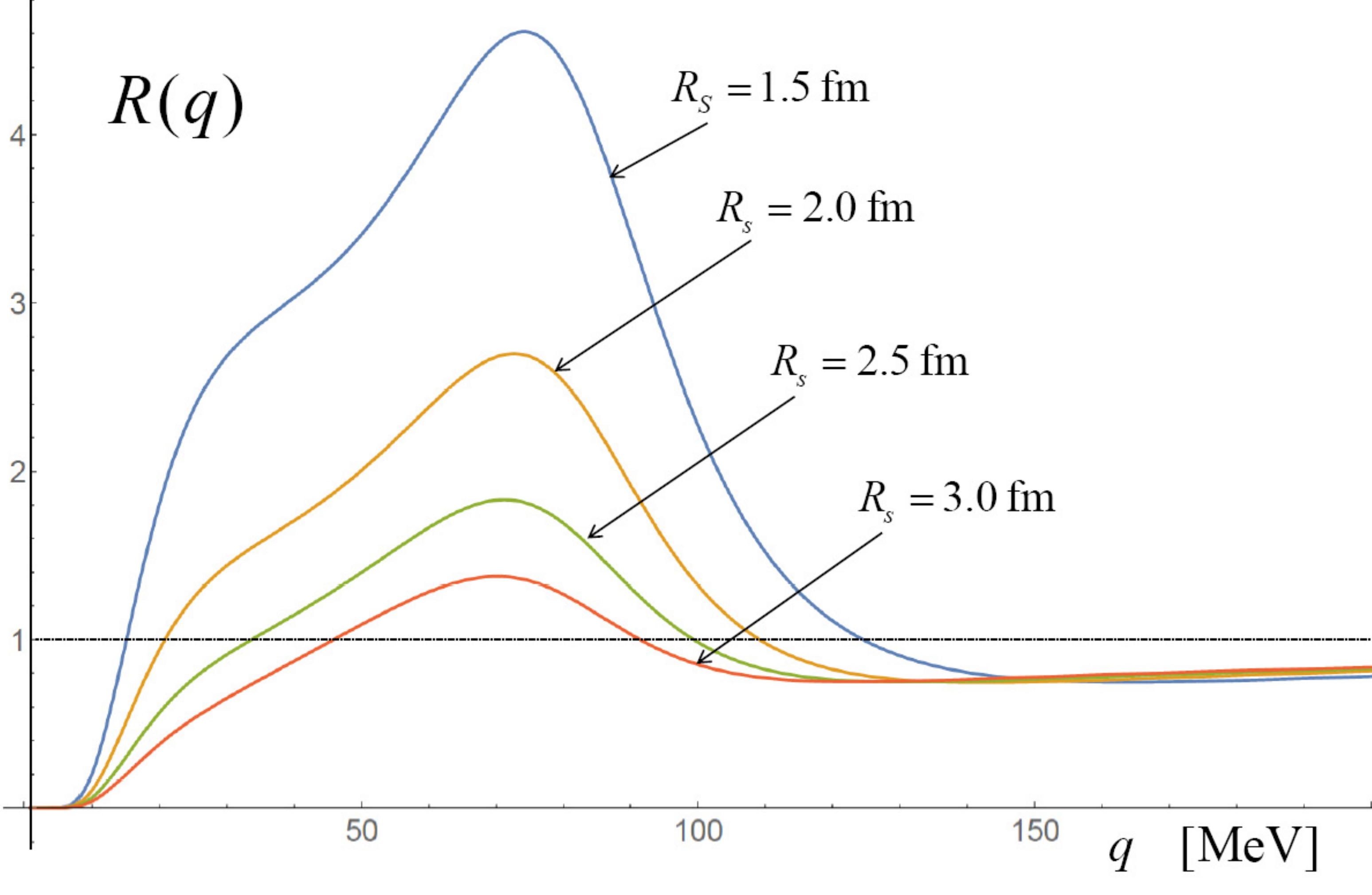}
\vspace{-2mm}
\caption{The spin-average $p\!-\!^3{\rm He}$ correlation function which takes into account the $s-$wave scattering, the resonance $^4{\rm Li}$ in the triplet channel and the Coulomb repulsion.}
\label{Fig-resonance-1-s-wave-Coulomb}
\end{figure}

In Fig.~\ref{Fig-resonance-1} we show the $p\!-\!^3{\rm He}$ correlation function which takes into account only the resonance interaction. The peak at $q=q_0=75.7$~MeV is well seen. Fig.~\ref{Fig-resonance-1-s-wave-Coulomb} shows the spin-average correlation function which takes into account the $s-$wave scattering, the resonance interaction and the Coulomb repulsion included by means of the Gamow factor (\ref{Gamow}). The resonance contributes only to the triplet correlation function because, as explained in Sec.~\ref{sec-4Li-4He}, the spin of $p$ and $^3{\rm He}$, which are constituents of $^4{\rm Li}$, equals unity. One observes that the  $s-$wave scattering and Coulomb repulsion strongly deform the resonance peak. 

A measurement of the $p\!-\!^3{\rm He}$ correlation function in relativistic heavy-ion collisions at LHC is difficult but possible. The correlation function was measured in $^{40}{\rm Ar}-$induced reactions on $^{197}{\rm Au}$ at the collision energy per nucleon of 60 MeV \cite{Pochodzalla:1987zz}. It was also measured in relativistic ${\rm Au}\!-\!{\rm Pt}$ collisions at AGS \cite{Armstrong:2001mr}.

One asks whether the $p\!-\!^3{\rm He}$ correlation function is sensitive enough to the change of the source radius from $R_s$ to $\sqrt{3/2}\,R_s$ that allows one to judge about the origin of $^3{\rm He}$: whether the nuclides are directly emitted from the source or they are formed due to final-state interactions. The answer obviously depends on accuracy of the $p\!-\!^3{\rm He}$ correlation function to be experimentally measured but Fig.~\ref{Fig-resonance-1-s-wave-Coulomb} suggests that it will be a difficult task. The problem is that  the source radius $R_s$ must be inferred from experimental data together with the parameter $\lambda_R$ which controls the resonance strength.

\subsection{Yield of $^4{\rm Li}$}
\label{sec-yield-Li}

As discussed in the previous section, the resonance peak of the $p\!-\!^3{\rm He}$ correlation function is distorted by the Coulomb repulsion and $s-$wave scattering. So, it is not obvious how to infer the resonance yield from the distribution of the $p\!-\!^3{\rm He}$  pairs.  

To derive the yield of $^4{\rm Li}$, we start with the formula (\ref{fun-corr-p-3He-elem}), which is written as 
\be
\label{yield-2}
\frac{dN_{p{^3{\rm He}}}}{d^3q  d^3P} = \mathcal{R}({\bf q}) \, 
\frac{dN_p}{d^3p_p}  \frac{dN_{^3{\rm He}}}{d^3p_{^3{\rm He}}}\bigg|_{{\bf q}=0} ,
\ee
where ${\bf P} \equiv {\bf p}_p + {\bf p}_{^3{\rm He}}$. The correlation function strongly depends on ${\bf q}$ but the dependence of the product $\frac{dN_p}{d^3p_p}  \frac{dN_{^3{\rm He}}}{d^3p_{^3{\rm He}}}$ on ${\bf q}$ is rather weak in the momentum domain of the resonance. So, it is taken at ${\bf q}=0$. 

To get the yield of $^4{\rm Li}$, one should sum up the number of correlated $p\!-\!^3{\rm He}$ pairs within the resonance peak. However, the peak is deformed by the Coulomb repulsion and $s-$wave scattering. So, we suggest to fit an experimentally obtained correlation function with the theoretical formula (\ref{fun-corr-gen-final}) where $\lambda_R$, which enters the amplitude (\ref{B-W-amplitude}) to control a strength of the resonance, is treated as a free parameter. Then, the contribution from the resonance can be disentangled. 

\begin{figure}[t]
\centering
\vspace{1mm}
\includegraphics[scale=0.24]{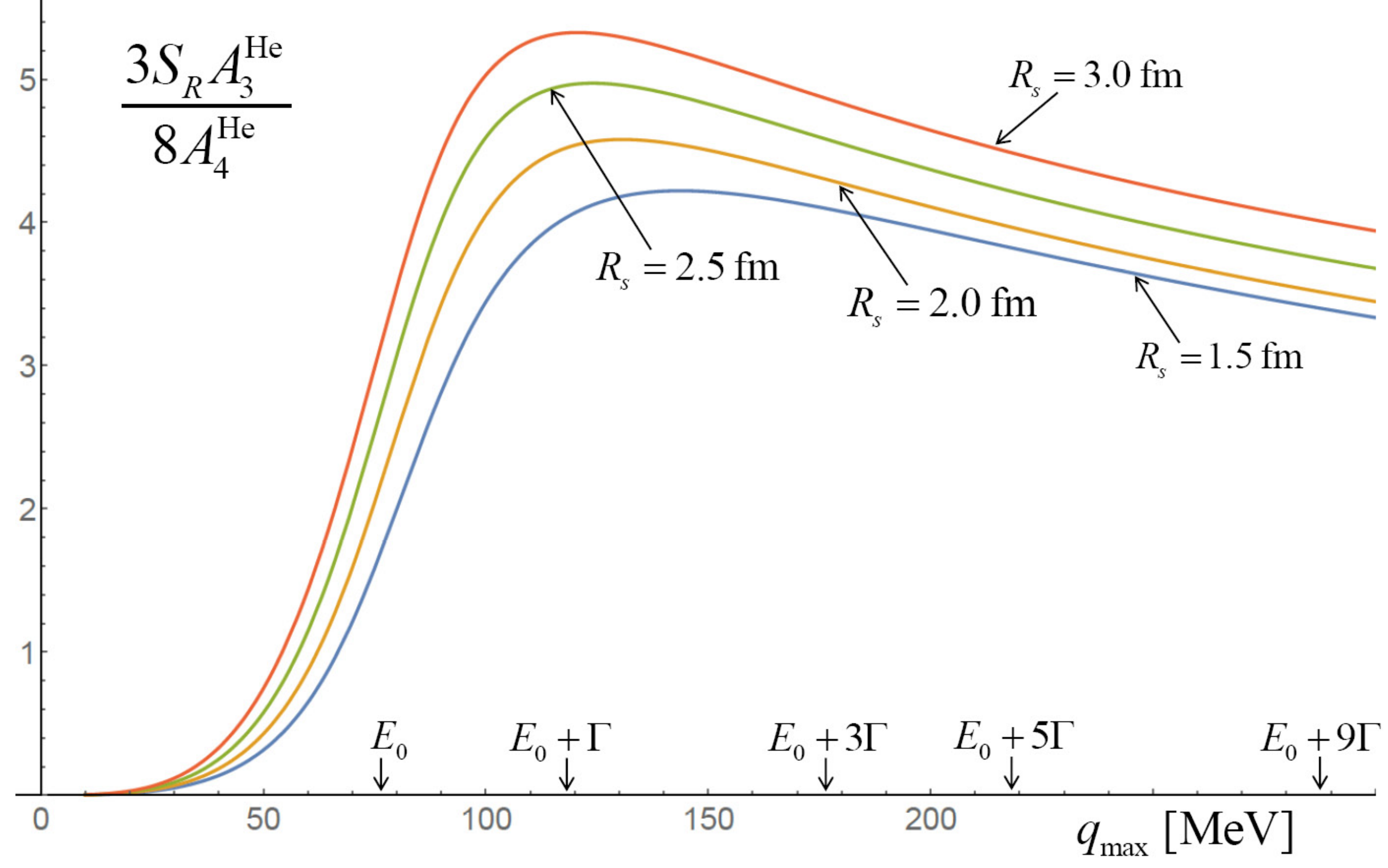}
\vspace{-1mm}
\caption{The ratio of $^4{\rm Li}$ to $^4{\rm He}$ yields as a function of $q_{\rm max}$ for four values of $R_s$.}
\label{Fig-ratio-qmax}
\end{figure}

Denoting the correlation function shown in Fig.~\ref{Fig-resonance-1}, which differs from unity solely due to the resonance interaction, as $\mathcal{R}_R({\bf q})$, the yield of $^4{\rm Li}$ of the momentum ${\bf P}$ equals 
\be
\label{yield-final}
\frac{dN_{{^4{\rm Li}}}}{d^3P} = \frac{3}{4} \, S_R 
\frac{dN_p}{d^3p_p}  \frac{dN_{^3{\rm He}}}{d^3p_{^3{\rm He}}}\bigg|_{{\bf q}=0} ,
\ee
where the factor $3/4$ takes into account that  $^4{\rm Li}$ is produced only in the triplet channel and
\be
\label{sR-sR-qmax} 
S_R \equiv 4\pi \int_0^{q_{\rm max}} \Big(\mathcal{R}_R(q) - 1\Big) \,q^2 dq.
\ee
Since the source function is assumed to be isotropic and the correlation function depends on ${\bf q}$ only through $q$, the trivial angular integration has been performed in Eq.~(\ref{sR-sR-qmax}). The  upper limit of the integral (\ref{sR-sR-qmax}) should be chosen in a such way that the integral covers the resonance peak centered at $q_0 \approx 76~~{\rm MeV}/c$. 

To get the ratio of the yields of $^4{\rm Li}$ to $^4{\rm He}$ one has to express the yields of $^3{\rm He}$ and of protons through the yields of nucleons. Keeping in mind the coalescence formula (\ref{A-mom-dis}), Eq.~(\ref{yield-final}) is written as 
\be
\label{yield-final-N}
\frac{dN_{{^4{\rm Li}}}}{d^3P} = \frac{3}{8} \, S_R \, {\cal A}_3
\bigg(\frac{dN_N}{d^3p_N} \bigg)^4 ,
\ee
where the additional factor $1/2$ takes into account that half of nucleons are protons. Consequently, the ratio of $^4{\rm Li}$ to $^4{\rm He}$ yields equals
\be
\frac{ ^4{\rm Li}}{ ^4{\rm He}} = \frac{3 \, S_R \, {\cal A}_3}{8{\cal A}_4^{\rm He}} .
\ee

In Fig.~\ref{Fig-ratio-qmax} we show the ratio as a function of $q_{\rm max}$ for $\lambda_R = 1$ and four values of $R_s$. There are indicted the values of relative momenta of $^3{\rm He}$ and $p$ (in the center-of-mass frame) which correspond to the energy of the resonance peak $E_0$, to $E_0 + \Gamma$, to $E_0 + 3\Gamma$, etc. The integral (\ref{sR-sR-qmax}) is seen to change rather slowly for $q_{\rm max}$ bigger than, say, 150 MeV but it is not clear whether the integral saturates when $q_{\rm max} \to \infty$. As observed in Ref.~\cite{Maj:2004tb} and further studied in \cite{Maj:2019hyy}, the analogous integrals of correlation functions usually diverge as $q_{\rm max} \to \infty$ because the correlation functions tend to unity as $q^{-3}$ or slower. However, it is not physically reasonable to extend the integral (\ref{sR-sR-qmax}) to a value of $q_{\rm max}$ higher than, say, $q_{\rm max} = 177$ MeV which corresponds to $E_0 + 3\Gamma$. The value of $S_R$ does not change very much when $q_{\rm max}$ is increased from 177 MeV to 286 MeV with the latter value corresponding to $E_0 + 9\Gamma$. Finally we observe that for $\lambda_R = 1$ and $q_{\rm max} \approx 150$ MeV the ratio varies between 4 and 5.5.  

At the end we note that the yield of $^4{\rm Li}$ was measured in ${\rm Au}\!-\!{\rm Pt}$ collisions at AGS \cite{Armstrong:2001mr} but it is unclear how the problem of distorted resonance peak was resolved. 

\section{Closing remarks}
\label{sec-conclusions}

It is truly surprising result that production of light nuclei at the highest accessible energies of heavy-ion collisions is equally well described by two completely different models. Our objective was to broadly present the problem and to consider proposals to falsify one of the models. The measurements which are proposed are challenging but possible. Hopefully, we will learn weather the ideas discussed here are actually useful not in a remote future. 

\section*{Acknowledgments}

This work was partially supported by the National Science Centre, Poland under grant 2018/29/B/ST2/00646. 



\begin{thebibliography}{99}

\bibitem{Adler:2001uy} 
C.~Adler {\it et al.} [STAR Collaboration],
Phys.\ Rev.\ Lett.\  {\bf 87}, 262301 (2001).

\bibitem{Agakishiev:2011ib} 
H.~Agakishiev {\it et al.} [STAR Collaboration],
Nature {\bf 473}, 353 (2011), 
Erratum: [Nature {\bf 475}, 412 (2011)].

\bibitem{Adam:2015vda} 
J.~Adam {\it et al.} [ALICE Collaboration],
Phys.\ Rev.\ C {\bf 93}, 024917 (2016).

\bibitem{Acharya:2017fvb} 
S.~Acharya {\it et al.} [ALICE Collaboration],
Phys.\ Rev.\ C {\bf 97}, 024615 (2018).

\bibitem{Acharya:2017bso} 
S.~Acharya {\it et al.} [ALICE Collaboration],
Nucl.\ Phys.\ A {\bf 971}, 1 (2018).

\bibitem{Abelev:2010rv} 
B.~I.~Abelev {\it et al.} [STAR Collaboration],
Science {\bf 328}, 58 (2010).

\bibitem{Adam:2015yta} 
J.~Adam {\it et al.} [ALICE Collaboration],
Phys.\ Lett.\ B {\bf 754}, 360 (2016).

\bibitem{Butler:1963pp} 
S.~T.~Butler and C.~A.~Pearson,
Phys.\ Rev.\  {\bf 129}, 836 (1963).

\bibitem{Schwarzschild:1963zz} 
A.~Schwarzschild and C.~Zupancic,
Phys.\ Rev.\  {\bf 129}, 854 (1963).

\bibitem{Sun:2015ulc} 
K.~J.~Sun and L.~W.~Chen,
Phys.\ Rev.\ C {\bf 93}, 064909 (2016).

\bibitem{Sun:2017ooe} 
K.~J.~Sun and L.~W.~Chen,
Phys.\ Rev.\ C {\bf 95}, 044905 (2017).

\bibitem{Zhu:2015voa} 
L.~Zhu, C.~M.~Ko and X.~Yin,
Phys.\ Rev.\ C {\bf 92}, 064911 (2015).

\bibitem{Zhu:2017zlb} 
L.~Zhu, H.~Zheng, C.~Ming Ko and Y.~Sun,
Eur.\ Phys.\ J.\ A {\bf 54}, 175 (2018).

\bibitem{Wang:2017smh} 
R.~q.~Wang, J.~Song, G.~Li and F.~l.~Shao,
Chin.\ Phys.\ C {\bf 43}, 024101 (2019).

\bibitem{BraunMunzinger:2003zd}
 P.~Braun-Munzinger, K.~Redlich and J.~Stachel,
in {\it Quark gluon plasma 3}, edited by  R.C. Hwa and X.N. Wang, World Scientific, Singapore, 2004, pp 491-599.

\bibitem{Andronic:2010qu}
A.~Andronic, P.~Braun-Munzinger, J.~Stachel and H.~Stocker,
Phys.\ Lett.\ B {\bf 697}, 203 (2011).

\bibitem{Cleymans:2011pe}
J.~Cleymans, S.~Kabana, I.~Kraus, H.~Oeschler, K.~Redlich and N.~Sharma,
Phys.\ Rev.\ C {\bf 84}, 054916 (2011).

\bibitem{Andronic:2017pug} 
A.~Andronic, P.~Braun-Munzinger, K.~Redlich and J.~Stachel,
Nature {\bf 561}, 321 (2018).

\bibitem{Mrowczynski:2016xqm} 
St.~Mr\'owczy\'nski,
Acta Phys.\ Polon.\ B {\bf 48}, 707 (2017).

\bibitem{Bazak:2018hgl} 
S.~Bazak and St.~Mr\'owczy\'nski,
Mod.\ Phys.\ Lett.\ A {\bf 33}, 1850142 (2018).

\bibitem{Mrowczynski:2019yrr} 
St.~Mr\'owczy\'nski and P.~S\l o\'n,
Acta Phys.\ Polon.\ B {\bf 51}, 1739  (2020).

\bibitem{Bazak:2020wjn} 
S.~Bazak and St.~Mr\'owczy\'nski,
Eur. Phys. J. A {\bf 56}, 193 (2020).

\bibitem{Bellini:2018epz} 
F.~Bellini and A.~P.~Kalweit,
Phys.\ Rev.\ C {\bf 99}, 054905 (2019).

\bibitem{Oliinychenko:2018ugs} 
D.~Oliinychenko, L.~G.~Pang, H.~Elfner and V.~Koch,
Phys.\ Rev.\ C {\bf 99}, 044907 (2019).

\bibitem{Xu:2018jff} 
X.~Xu and R.~Rapp,
Eur.\ Phys.\ J.\ A {\bf 55}, 68 (2019).

\bibitem{Bugaev:2018klr} 
K.~A.~Bugaev {\it et al.},
J.\ Phys.\ Conf.\ Ser.\  {\bf 1390}, 012038 (2019).

\bibitem{Bugaev:2020sgz}
K.~A.~Bugaev {\it et al.},
arXiv:2005.01555 [nucl-th].

\bibitem{Sun:2018mqq} 
K.~J.~Sun, C.~M.~Ko and B.~D\" onigus,
Phys. Lett B {\bf 792}, 132 (2019).

\bibitem{Vovchenko:2019aoz} 
V.~Vovchenko, K.~Gallmeister, J.~Schaffner-Bielich and C.~Greiner,
Phys.\ Lett.\ B {\bf 800}, 135131 (2020).

\bibitem{Vovchenko:2020dmv} 
V.~Vovchenko, B.~D\"onigus, B.~Kardan, M.~Lorenz and H.~Stoecker,
Phys. Lett. B \textbf{809}, 135746 (2020).

\bibitem{Cai:2019jtk} 
Y.~Cai, T.~D.~Cohen, B.~A.~Gelman and Y.~Yamauchi,
Phys.\ Rev.\ C {\bf 100}, 024911 (2019).

\bibitem{DasGupta:1981xx}
S.~Das Gupta and A.~Mekjian,
Phys. Rept. \textbf{72}, 131 (1981).

\bibitem{Sombun:2018yqh}
S.~Sombun, K.~Tomuang, A.~Limphirat, P.~Hillmann, C.~Herold, J.~Steinheimer, Y.~Yan and M.~Bleicher,
Phys. Rev. C \textbf{99}, 014901 (2019).

\bibitem{Sato:1981ez} 
H.~Sato and K.~Yazaki,
Phys.\ Lett.\ B {\bf 98}, 153 (1981).

\bibitem{Gyulassy:1982pe} 
M.~Gyulassy, K.~Frankel and E.~A.~Remler,
Nucl.\ Phys.\ A {\bf 402}, 596 (1983).

\bibitem{Mrowczynski:1987} 
St. Mr\' owczy\' nski, J. Phys. G {\bf 13}, 1089 (1987).

\bibitem{Lyuboshitz:1988} 
V.L. Lyuboshitz, Sov. J. Nucl. Phys. {\bf 48}, 956 (1988) [Yad. Fiz. {\bf 48}, 1501 (1988)].

\bibitem{Mrowczynski:1992gc} 
St.~Mr\'owczy\'nski,
Phys.\ Lett.\ B {\bf 277}, 43 (1992).

\bibitem{Pais:2019shp}
H.~Pais, F.~Gulminelli, C.~Provid\^encia and G.~R\"opke,
Phys. Rev. C \textbf{99}, 055806 (2019). 

\bibitem{Ropke:2020peo}
G.~R\"opke,
Phys. Rev. C \textbf{101}, 064310 (2020). 

\bibitem{Mrowczynski:1994rn} 
St.~Mr\'owczy\'nski,
Phys.\ Lett.\ B {\bf 345}, 393 (1995).

\bibitem{Maj:2004tb} 
R.~Maj and St.~Mr\'owczy\'nski,
Phys.\ Rev.\ C {\bf 71}, 044905 (2005).

\bibitem{Maj:2019hyy} 
R.~Maj and St.~Mr\'owczy\'nski,
Phys.\ Rev.\ C {\bf 101}, 014901 (2020).

\bibitem{Brown:1997ku} 
D.~A.~Brown and P.~Danielewicz,
Phys.\ Lett.\ B {\bf 398}, 252 (1997).

\bibitem{Alt:2008aa} 
C.~Alt {\it et al.} [NA49 Collaboration],
Phys.\ Lett.\ B {\bf 685}, 41 (2010).

\bibitem{Maj:2009ue} 
R.~Maj and St.~Mr\'owczy\'nski,
Phys.\ Rev.\ C {\bf 80}, 034907 (2009).

\bibitem{PBM-2015}
P.~Braun-Munzinger, B.~D\"onigus and N.~L\"oher, CERN Courier, August 2015.

\bibitem{Angeli:2013epw} 
I.~Angeli and K.~P.~Marinova,
Atom.\ Data Nucl.\ Data Tabl.\ {\bf 99}, 69 (2013).

\bibitem{Adam:2015vna} 
J.~Adam {\it et al.} [ALICE Collaboration],
Phys.\ Rev.\ C {\bf 93}, 024905 (2016).

\bibitem{Adam:2015vja} 
J.~Adam {\it et al.} [ALICE Collaboration],
Phys.\ Rev.\ C {\bf 92}, no. 5, 054908 (2015).

\bibitem{Bergstrom:1979gpv} 
J.~C.~Bergstrom,
Nucl.\ Phys.\ A {\bf 327}, 458 (1979).

\bibitem{Acharya:2018gyz} 
S.~Acharya {\it et al.} [ALICE Collaboration],
Phys.\ Rev.\ C {\bf 99}, 024001 (2019).

\bibitem{Ardouin:1996nc}
D.~Ardouin,
Int. J. Mod. Phys. E \textbf{6}, 391 (1997).

\bibitem{Laura} L.~Fabbietti, private communication.

\bibitem{NNDC}
National Nuclear Data Center, Chart of Nuclides, http://www.nndc.bnl.gov.

\bibitem{Tilley:1992zz} 
D.~R.~Tilley, H.~R.~Weller and G.~M.~Hale,
Nucl.\ Phys.\ A {\bf 541}, 1 (1992).

\bibitem{Koonin:1977fh} 
S.~E.~Koonin,
Phys.\ Lett.\  {\bf 70B}, 43 (1977).

\bibitem{Lednicky:1981su} 
R.~Lednicky and V.~L.~Lyuboshitz,
Sov.\ J.\ Nucl.\ Phys.\  {\bf 35}, 770 (1982)  [Yad.\ Fiz.\  {\bf 35}, 1316 (1981)].

\bibitem{Gmitro:1986ay} 
M.~Gmitro, J.~Kvasil, R.~Lednicky and V.~L.~Lyuboshitz,
Czech.\ J.\ Phys.\ B {\bf 36}, 1281 (1986).

\bibitem{Black:1999duc} 
T.~C.~Black, H.~J.~Karwowski, E.~J.~Ludwig, A.~Kievsky, S.~Rosati and M.~Viviani,
Phys.\ Lett.\ B {\bf 471}, 103 (1999).

\bibitem{Daniels:2010af} 
T.~V.~Daniels, C.~W.~Arnold, J.~M.~Cesaratto, T.~B.~Clegg, A.~H.~Couture, H.~J.~Karwowski and T.~Katabuchi,
Phys.\ Rev.\ C {\bf 82}, 034002 (2010).

\bibitem{Landau-Lifshitz-1988}
L.D.~Landau and E.M.~Lifshitz, {\it Quantum Mechanics} (Pergamon Press, Oxford, 1988).

\bibitem{Pochodzalla:1987zz} 
J.~Pochodzalla {\it et al.},
Phys.\ Rev.\ C {\bf 35}, 1695 (1987).

\bibitem{Armstrong:2001mr} 
T.~A.~Armstrong {\it et al.},
Phys.\ Rev.\ C {\bf 65}, 014906 (2002).

\end{thebibliography}
\end{document}